\documentclass{aa}
\usepackage{graphicx}
\usepackage{txfonts}
\usepackage{amsmath}
\usepackage{xcolor}
\usepackage{mhchem}
\usepackage{hyperref}

\begin{document}

   \title{Mineral snowflakes on exoplanets and brown dwarfs}

   \subtitle{Effects of micro-porosity, size distributions, and particle shape}

   \author{D. Samra \inst{1,2}
          \and
          Ch. Helling \inst{1,2,3}
          \and
          M. Min \inst{3,4}
          }

   \institute{Centre for Exoplanet Science, University of St Andrews, North Haugh, St Andrews, KY169SS, UK\\
             \email{dbss3@st-andrews.ac.uk}
         \and
             SUPA, School of Physics \& Astronomy, University of St Andrews, North Haugh, St Andrews, KY169SS, UK
         \and
         SRON Netherlands Institute for Space Research, Sorbonnelaan 2, 3584 CA Utrecht, NL
         \and
         Astronomical Institute Anton Pannekoek, University of Amsterdam, Science Park 904, 1098 XH Amsterdam, The Netherlands}
   \date{Received 22 January 2020; accepted 17 April 2020}

 
  \abstract
   {Exoplanet atmosphere characterisation has become an important tool in understanding exoplanet formation, evolution, and it also is a window into potential habitability. However, clouds remain a key challenge for characterisation: upcoming space telescopes (e.g.  the James Webb Space Telescope, {\sc JWST}, and the Atmospheric Remote-sensing Infrared Exoplanet Large-survey, {\sc ARIEL}) and ground-based high-resolution spectrographs (e.g. the next-generation CRyogenic high-resolution InfraRed Echelle Spectrograph, {\sc CRIRES+}) will produce data requiring detailed understanding of cloud formation and cloud effects for a variety of exoplanets and brown dwarfs.}
   {We aim to understand how the micro-porosity of cloud particles affects the cloud structure, particle size, and material composition on exoplanets and brown dwarfs. We further examine the spectroscopic effects of micro-porous particles, the particle size distribution, and non-spherical cloud particles.}
   {We expanded our kinetic non-equilibrium cloud formation model to study the effect of micro-porosity on the cloud structure using prescribed 1D ($T_{\rm gas}$-$p_{\rm gas}$) profiles from the {\sc Drift-Phoenix} model atmosphere grid. We applied the effective medium theory and the Mie theory to model the spectroscopic properties of cloud particles with micro-porosity and a derived particle size distribution. In addition, we used a statistical distribution of hollow spheres to represent the effects of non-spherical cloud particles.}
    {Highly micro-porous cloud particles (90\% vacuum) have a larger surface area, enabling efficient bulk growth higher in the atmosphere than for compact particles. Increases in single scattering albedo and cross-sectional area for these mineral snowflakes cause the cloud deck to become optically thin only at a wavelength of $\sim 100\,{\rm \mu m}$ instead of at the $\sim 20\,{\rm \mu m}$ for compact cloud particles. A significant enhancement in albedo is also seen when cloud particles occur with a locally changing Gaussian size distribution. Non-spherical particles increase the opacity of silicate spectral features, which further increases the wavelength at which the clouds become optically thin.}
    {Retrievals of cloud properties, particularly particle size and mass of clouds, are biased by the assumption of compact spherical particles. The {\sc JWST} {\sc } mid-infrared instrument (MIRI) will be sensitive to signatures of micro-porous and non-spherical cloud particles based on the wavelength at which clouds are optically thin. Details of spectral features are also dependent on particle shape, and greater care must be taken in modelling clouds as observational data improves.}


   \maketitle

\section{Introduction}
\label{sec:Intro}
The first detection of an exoplanet atmosphere showed diminished sodium spectral lines for the hot Jupiter HD 209458 b, which is suggestive of clouds in the atmosphere \citep{Charbonneau2002}. This was in line with the expectation that interpreting the transmission spectra of exoplanets would require careful consideration of condensates \citep{Seager2000}. Similarly, the importance of clouds had previously been recognised for brown dwarf atmospheres \citep{Lunine1986,Tsuji1996a,Tsuji1996b}. Since the first detection of truncated spectral features, spectra indicative of cloud particles (also called {\it \textup{aerosols}}) have been found for a number of exoplanets (e.g. \citet{Benneke2012,Crossfield2013,Deming2013,Kreidberg2014,Sing2016,Kreidberg2018}). As clouds became established as a regular feature of sub-stellar atmospheres, many research groups developed models using very different inspirations: based on terrestrial cloud formation \citep{Ackerman2001,Cooper2003}, from the point of view of planetary science \citep{Rossow1978,Marley1999}, of asymptotic giant branch (AGB) stars \citep{Helling2001,Woitke2003,Woitke2004}, and based on practical considerations \citep{Tsuji1996a,Tsuji1996b,Allard2001}. Detailed comparisons of all these models can be found in \cite{Helling2008c} and \cite{Charnay2018}.

Cloud particle size distributions in the literature often use assumed distribution forms, where the value of the parameters describing the distribution are derived from observational data (e.g. the model of \citet{Ackerman2001}, which uses a log-normal distribution). One method for calculating the particle size distribution without imposing a form is the binning method, which has been used to model the coagulation of grains in a protoplanetary disc \citep{Dullemond2005,Birnstiel2010}. The binning method has also been applied more recently to clouds on hot Jupiters \citep{Powell2018} and super-Earths \citep{Gao2018b,Gao2018a}, and it was used for photochemical hazes \citep{Kawashima2018}. Alternatively, cloud formation in an atmosphere can be modelled using the moment method, which is computationally fast, and a size distribution can be reconstructed based on this \citep{Deuflhard1989,Krueger1995}, ideally using many moments to represent the local particle size distribution.

Because of practical considerations, retrievals of exoplanet atmospheres try to limit the parameter space as much as possible, often assuming particle sizes a priori or modelling clouds as a grey cloud deck (e.g. \citet{Madhusudhan2011,Benneke2012,Kreidberg2014}). Recently, attempts have been made to include the Mie theory to fully calculate properties of the clouds in retrieval, see \citet{Benneke2019a}, but in a later study, the authors also found that a grey cloud model for the habitable zone planet K2-18b \citep{Benneke2019b} was a better fit to the data. This highlights the limitations of the low-resolution spectra that are currently available. Future instruments that will be capable of providing higher resolution spectra, such as  the next-generation CRyogenic high-resolution InfraRed Echelle Spectrograph ({\sc CRIRES+)} \citep{Follert2014}, or observations with a higher signal-to-noise ratio such as are expected from the James Webb Space Telescope ({\sc JWST)} \citep{Gardner2006} and the Atmospheric Remote-sensing Infrared Exoplanet Large-survey {\sc (ARIEL)} \citep{Tinetti2018}, will require a better understanding of the effects of simplifying assumptions in models on the distribution and optical effects of clouds.

Micro-porosity is the porosity arising from the organisation of the condensate monomers (e.g. \ce{Mg2SiO4} in \ce{Mg2SiO4}[s]) within a cloud particle during growth. This is different from the porosity that can be used to characterise aggregates that originate from particle-particle collision processes (coagulation, e.g. \cite{Dominik1997,Blum2000}), which we do not consider here. On Earth, the material density of water ice is dependent on the ambient temperature at formation. Snowflakes are known to form many types of crystal structures that can be up to 84\% porous for millimetre-sized cloud particles when compared to ice material density \citep{Hales2005}, leading to the possibility of altitude-dependent porosity in terrestrial snow clouds. Earth-like exoplanets, mini-Neptunes, and T-type brown dwarfs may form water clouds, composed of liquid or solid particles, but warmer planets and brown dwarfs of L-type and later have been shown to form cloud particles made of a mix of materials that is dominated by Mg, Si, Fe, and O and to a lesser extent by Ti, Al, K and other elements (e.g. \citet{Witte2009,Lee2015a,Helling2019b}). There are many ways in which this micro-porosity might be incorporated into mineral cloud particles, for example lattice faults at the interfaces between two different condensation species owing to the different lattice structures. Even for homogeneous growth, single species often have multiple crystal structures \citep{Sood2013}, which can also generate lattice faults at their interfaces. For example, the \ce{TiO2}[s] rutile and anatase forms are both stable at atmospheric pressures for temperatures greater than 1100 K \citep{Jung2001,Hanaor2010}. Additionally, within crystal structures, there are many known types of defect that might further decrease material density (e.g. Schottky defects in \ce{TiO2}[s] and \ce{MgO}[s] crystals \citep{Mntrey2004}). Furthermore, these cloud particles not only change their material composition when falling through the atmosphere, but their particle sizes will also change such that the largest cloud particles are forming the innermost part of the cloud, which often sits deep inside the optically thick part of the atmosphere. Because cloud particles made of a mix of many thermally stable materials fall into warmer atmospheric regions, the low-temperature materials (such as SiO[s], MgSiO3[s]) become thermally unstable, they evaporate and leave behind a skeleton made of high-temperature materials (such as Fe[s], TiO2[s], Al2O3[s]). Whilst this may be a source of micro-porosity of cloud particles, \citet{Juncher2017} noted that this may also lead to a reduction in micro-porosity because the structural integrity of the particle is weakened and dangling structures break off. These micro-porous mineral cloud particles we call `mineral snowflakes'.

The aim of this paper is to assess how simplifying assumptions about cloud particles, such as sphericity, homodispersity, and compactness, affect the spectral properties of clouds. We investigate this in the framework of our kinetic non-equilibrium cloud formation model. We confine the study to atmospheric models typical of brown dwarfs and Jupiter-size gas-giant exoplanets, but we anticipate a wider applicability to other exoplanets such as mini-Neptunes, super-Earths, and lava worlds. 

In Section~\ref{sec:Cloud_form} we briefly summarise our approach to modelling cloud formation, how we model micro-porosity, the cloud particle size distribution function, and our approach to modelling the non-sphericity of cloud particles as part of our opacity calculations. Section~\ref{sec:Por_effects} shows the effects of micro-porous particles, first for an atmosphere model of a typical warm gas giant, and then for a range of effective temperatures and surface gravities. In Section~\ref{sec:Sizedist_effects} we investigate the effects on albedo of a size distribution of cloud particles. Section~\ref{sec:DHS_optical_effects} describes the optical effects of non-spherical cloud particles. Section~\ref{sec:Coupled_effects} shows the combined optical effects of the three deviations from compact, monodisperse, and spherical particles.

\section{Cloud formation model}
\label{sec:Cloud_form}
The principal problem of cloud formation is the efficient conversion of a gas into bulk cloud particles. This requires favourable local thermochemical conditions and sufficient time for the necessary reactions to occur. Our kinetic non-equilibrium model with consistent gas-phase element conservation describes the formation of cloud particles by nucleation, bulk growth, evaporation, gravitational settling, and turbulent mixing. We briefly describe the key aspects here. A full explanation can be found in \citet{Woitke2003}, \citet{Woitke2004}, \citet{Helling2006a}, \citet{Helling2006b}, \citet{Helling2008a}, \citet{Helling2013}, and \citet{Helling2019}.

\textit{\textup{Nucleation}} describes the formation of condensation seed particles. We applied the modified classical nucleation theory for homogeneous nucleation, which models cluster formation by gas-gas interactions. We calculated nucleation rates $J_{\rm i}$ for ${\rm i}\,=\,\ce{TiO2}{\rm \,[s]},\ce{SiO}{\rm \,[s]}$ as in \citet{Lee2015a}, and also for ${\rm i}\,=\,\ce{C}{\rm \,[s]}$. The sum of the three rates is the total nucleation rate $J_{*}\,=\,\Sigma_{\rm i} J_{\rm i}\,[\rm{cm^{3}\,s^{-1}}]$. As the only means of forming new cloud particles, this sets the number density of them in the atmosphere ($n_{\rm d}\,[{\rm cm^{-3}}]$). For a material to form clusters directly from the gas phase, the gas has to be highly supersaturated (supersaturation ratio $S\,>>\,1$), and for this to occur, the gas must be much cooler than the temperature required for the thermal stability of a material \citep{Helling2019,Goeres1996}.

\textit{\textup{Bulk growth}} is the net deposition of material from the gas phase onto the surface of existing cloud particles. For this to happen, the material must be thermally stable as a condensate \citep{Helling2013}, that is, the rate of condensate evaporation is lower than the deposition rate (equivalent to $S\,>\,1$). Condensation onto the surface of a cloud particle is significantly more energy efficient than nucleation, and therefore only mild supersaturation is required. We considered the formation of 15 bulk materials (s~=~\ce{TiO2}[s], \ce{Mg2SiO4}[s], \ce{MgSiO3}[s], MgO[s], SiO[s], \ce{SiO2}[s], Fe[s], FeO[s], FeS[s], \ce{Fe2O3}[s], \ce{Fe2SiO4}[s], \ce{Al2O3}[s], \ce{CaTiO3}[s], \ce{CaSiO3}[s], and C[s]) by 126 gas-surface reactions (see \citet{Helling2019a}). Nucleating species were also considered for surface growth reactions because both processes deplete the gaseous elemental abundances of the forming elements.

\textit{\textup{Evaporation}} is the inverse process to bulk growth and occurs when cloud particles containing a condensed material reach the point in the atmosphere where that particular material is no longer thermally stable ($S\,<\,1$). In these regions the deposition rate of the material is therefore exceeded by the evaporation rate, and the material evaporates from the cloud particles. 

\textit{\textup{Gravitational settling}} occurs when the frictional forces exerted on cloud particles by the gas are no longer sufficient to couple them to the gas, and thus the particles fall out of that layer of the atmosphere. The velocities at which the cloud particles settle in the atmosphere with respect to the gas quickly reach an equilibrium value called the `drift velocity'. For a subsonic free molecular flow ($Kn\,<<\,1$), this is given by \citet{Woitke2003} as

\begin{equation}
    \langle\,\mathring{v}_{\rm{dr}}\,\rangle\,=\,\frac{\sqrt{\pi}}{2}\frac{\varg \rho_{\rm s} \langle\,a\,\rangle}{\rho c_{\rm T}},
    \label{equ:driftv}
\end{equation}

which is dependent on the size of the cloud particle $\langle\,a\,\rangle$, gas density $\rho$ , and material density $\rho_{\rm s}$. Consequently, large cloud particles or cloud particles of higher density will settle faster. The inverse proportionality to gas density and mean thermal velocity $c_{\rm T}\,=\,\sqrt{2k_{\rm B} T/\overline{\mu}}$ means that cloud particles settle from higher in the atmosphere to a lower layer because gas density and temperature typically increase with depth in an atmosphere. For a sufficiently high drift velocity, the cloud particles settle faster than bulk growth can occur, resulting in the cloud particles remaining at a constant size and `raining out' of the atmosphere, which means that they rapidly fall towards lower atmospheric layers before they evaporate.

The combined processes of gravitational settling and evaporation naturally deplete the gas-phase element abundances of condensible material in regions where cloud particles initially form and conversely enriches it where cloud materials evaporate. For the gas phase, we assumed chemical equilibrium for 156 molecules, 16 atomic species, and various ionic species, consistently with the depletion in the cloud formation model. For gas species that are not affected by cloud formation, solar abundances were assumed \citep{Grevesse1992}. The abundances $\varepsilon_{\rm i}(z)$ of the elements i~=~\ce{O}, \ce{Ca}, \ce{S}, \ce{Al}, \ce{Fe}, \ce{Si}, \ce{Mg}, \ce{Ti}, and \ce{C} are affected by cloud formation processes. 

\textit{\textup{Turbulent mixing}} is necessary for sustained cloud formation; without replenishment of the upper atmosphere gas phase, cloud formation would cease because  the necessary condensable species would be depleted \citep{Woitke2004}. Large-scale convective motions provide such a process, but because this motion is inherently a 3D hydrodynamical effect, including it in 1D models requires parametrising it as a pseudo-diffusive term. We used a convective overshooting approach to parametrise the local turbulent gas-mixing timescale, according to Eq. 9 in \cite{Woitke2004} with $\beta = 1.0$. More recently, other mixing schemes have been tried, such as the diffusive approach. We refer to  the discussion of the details in \citet{Woitke2019}.

To model the cloud formation, we used the moment method that was first developed by \citet{Gail1988} for AGB star winds and later to calculate the interstellar medium (ISM) \citep{Krueger1995}. The moment method has the advantage of being computationally less expensive than the binning method \citep{Woitke2003}. Moments are defined by the integral

\begin{equation}
    \rho L_{j} = \int_{V_{\rm l}}^\infty f\left(V\right)V^{j/3} dV.
    \label{equ:moments}
\end{equation}

The lower integration boundary ($V_{\rm l}$) is the minimum cloud particle volume. We set this equivalent to 1000 \ce{TiO2}[s] monomers, which represents the size of our seed particles. The $j$th dust moment has units of $[{\rm cm}^{j}\rm{g^{-1}}],$ hence $\rho L_{j}$ has units of $[{\rm cm}^{j-3}]$. We solved the moment equations, Eq.~57 in \citet{Woitke2003}, for the formation of mixed-material cloud particles as developed in \citet{Woitke2003}, \citet{Woitke2004}, \citet{Helling2006a}, \citet{Helling2006b}, \citet{Helling2008a}, \citet{Helling2013}, and \citet{Helling2019}. From Eq.~\ref{equ:moments} it is clear that when $j\,=\,0,$ 

\begin{equation}
        n_{\rm d}\,=\,\rho L_0\,[\rm{cm^{-3}}].
        \label{equ:numden}
\end{equation}

Furthermore, the average cloud particle properties can be similarly computed from the moments, such as the mean particle size ($\langle\,a\,\rangle$),

\begin{equation}
    \langle\,a\,\rangle\,=\,\left(\frac{3}{4\pi}\right)^{1/3}\frac{L_1}{L_0}\,[\rm{cm}]
    \label{equ:meansize}
.\end{equation}

The solution of the moment equations provides us with information about the local mean particle sizes (Eq.~\ref{equ:meansize}), the number density of cloud particles (Eq.~\ref{equ:numden}), and also the material composition of the cloud particles. These all vary with height in the atmosphere because they depend on local thermodynamic conditions. 

\subsection{Opacities and Mie theory}
\label{subsec:Mie_theory}
Cloud particle opacities were calculated using the Mie theory \citep{Mie1908,Bohren1983}, as implemented by \citet{Wolf2004}. Refractive indices necessary for the Mie theory for the heterogeneous cloud particles were calculated using the effective medium theory with the Bruggeman mixing rule \citep{Bruggeman1935},

\begin{equation}
        \Sigma_{\rm s}\left(\frac{V_{\rm s}}{V_{\rm tot}}\right) \frac{\epsilon_{\rm s}-\epsilon_{\rm eff}}{\epsilon_{\rm s}+2\epsilon_{\rm eff}}\,=\,0,
        \label{equ:Brugg}
\end{equation}

where $\epsilon_{\rm s}$ is the dielectric constant for the individual condensate materials (s) that form the cloud particles. The refractive index of the condensate material is $m_{\rm s}\,=\,n_{\rm s} + ik_{\rm s}\,=\,\sqrt{\epsilon_{\rm s}}$. $V_{\rm s}/V_{\rm tot}$ is the volume fraction of cloud particles that is comprised of an individual condensate material. The desired effective dielectric constant of the cloud particles is $\epsilon_{\rm eff}$, found by solving Eq.~\ref{equ:Brugg} iteratively with the Newton-Raphson method. Extinction and scattering efficiency factors ($Q_{\rm ext} \text{ and }\,Q_{\rm sca}$ , respectively) calculated by the Mie theory are used to compute the single scattering albedo ($A_{\rm s}(\lambda)$) for the grains in a given layer. For a monodisperse distribution, this is

\begin{equation}
        A_{\rm S}(\lambda) = \frac{Q_{\rm sca}}{Q_{\rm ext}}.
        \label{equ:ssa}
\end{equation}

\subsection{Inputs and model setup}
\label{sec:Inputs}
\paragraph{Input profiles.} We used 1D {\sc Drift-Phoenix} atmosphere models \citep{Dehn2007,Helling2008a,Witte2009,Witte2011} as input for our cloud formation model. They provide ($T_{\rm gas}$-$p_{\rm gas}$) profiles calculated by consistently including cloud formation and its subsequent effects on the atmosphere opacity and element abundances. We used a subset of the models covering a grid of $T_{\rm eff}\,=\,1200,\,1800,\text{ and}\,2400\,{\rm K}$. We also used surface gravities $\log(\varg)\,=\,3.0$ and $5.0$, which is representative of gas giant exoplanets, and brown dwarfs and young gas giant exoplanets \citep{Witte2009}.

\paragraph{Optical constants.} We calculated cloud particle opacities for a log-linearly spaced grid of 100 wavelengths spanning the range $\lambda\,=\,0.1-1000\,\rm{\mu m}$. We used optical constants identical to those in \citet{Helling2019b} (reproduced in Table~\ref{tab:opt} and shown in Fig.~\ref{fig:appendix_refindex1}). Materials where the experimental data do not cover the full wavelength grid were treated as described in \citet{Lee2016}. Our data largely consist of values for amorphous materials. Whilst \citet{Kitzmann2018} have argued for the use of amorphous condensates, \citet{Helling2009} previously suggested that exoplanet cloud particles can be crystalline at temperatures exceeding $900\,{\rm K}$  because the thermal energy is sufficient to allow for lattice rearrangement within the cloud particle. The internal structure of the cloud particle material has implications for the spectra of ultra-hot Jupiters such as WASP-43b and WASP-121b at wavelengths that will be observable by the {\sc JWST} {\sc MIRI.} 

\subsection{Micro-porosity}
\label{subsec:Micro_por}
We modelled the micro-porosity of cloud particles by introducing an effective material density for each material ($\rho_{s}^{\rm eff}$). To do this, we modified the material density $\rho_{\rm s}$ by a `micro-porosity fraction' ($f_{\rm{\rm por}}$), which for this study was assumed constant for all condensate materials (s) and atmospheric layers. Hence 

\begin{equation}
    \rho_{\rm s}^{\rm eff} = \rho_{\rm s}\left(1-f_{\rm{\rm por}}\right).
    \label{equ:effden}
\end{equation}

For opacity calculations we incorporated the additional volume introduced by the micro-porosity factor as vacuum (using the effective medium theory), with the complex refractive index $m = 1$ (i.e. $n\,=\,1,\,k\,=\,0$). This simple approach enabled us to investigate the effect of micro-porosity on cloud optical and material properties for a variety of $f_{\rm{\rm por}}$ values. Previously, an effective medium approach was used for dust in protoplanetary discs \citep{Woitke2016} assuming a value of $f_{\rm por} = 0.25$.

\subsection{Cloud particle size distribution}
\label{subsec:Sizedist}
We used the results of our cloud formation models for the moments $L_{j}$ (Eq.~\ref{equ:moments}) to reconstruct a cloud particle size distribution function $f(a)$ through the related moments in radius-space $K_j$,

\begin{equation}
    K_j = \left(\frac{3}{4\pi}\right)^{j/3}\rho L_j = \int_{a_l}^\infty f\left(a\right)a^{j} da
    \label{equ:rad_moments}
.\end{equation}

For this study, we used the Gaussian distribution specified by three parameters: the mean particle size $\overline{a}\,[\rm{\mu m}]$, the standard deviation $\sigma\,[\rm{\mu m}]$, and the total number density of cloud particles $n_{\rm d}^{\rm Gauss}\,[{\rm cm^{-3}}]$ with the form

\begin{equation}
    f\left(a\right)\,=\,\frac{n_{\rm d}^{\rm Gauss}}{\sigma\sqrt{\pi}}\exp\left(-\frac{\left(a-\overline{a} \right)^2}{\sigma^2}\right).
    \label{equ:gaussian}
\end{equation}

Following the approach in \citet{Helling2008b}, we substituted Eq.~\ref{equ:gaussian} into Eq.~\ref{equ:rad_moments} and integrated across all cloud particle radii (extending the lower limit of the integral to $-\infty$). Thus we can write the parameters of the Gaussian distribution in terms of the moments in radius-space, $K_{j},\,j=0,1,2,3,4$, as

\begin{equation}
    n_{\rm d}^{\rm Gauss} = \frac{4K_{1}^{2}}{3K_{2}\pm\sqrt{9K_{2}^{2}-8K_{1}K_{3}}},
    \label{equ:gau_numden}
\end{equation}

\begin{equation}
    \overline{a} = \frac{K_{1}}{n_{\rm d}^{\rm Gauss}},
    \label{equ:gau_grainrad}
\end{equation}

\begin{equation}
    \sigma = \sqrt{\frac{2}{n_{\rm d}^{\rm Gauss}\left(K_{2}-n_{\rm d}^{\rm Gauss}\overline{a}^{2}\right)}}.
    \label{equ:gau_sigma}
\end{equation}

The derived distribution does not feedback on the moments, and therefore the cloud formation is unaffected by its assumed form. The derived parameters represent the information encapsulated by the moments, but the moments remain the same regardless of the reconstructed size distribution. The mean particle size of the distribution $\overline{a}$ (Eq.~\ref{equ:gau_grainrad}) is distinct from the average particle size from the moments $\langle\,a\,\rangle$ (Eq.~\ref{equ:meansize}). Combining Eqs.~\ref{equ:gau_grainrad} and \ref{equ:rad_moments}, we can write 

\begin{equation}
        \overline{a} = \left( \frac{3}{4\pi} \right)^{1/3} \frac{\rho L_1}{n_{\rm d}^{\rm Gauss}}.
        \label{equ:alt_gau_grainrad}
\end{equation}

Comparing this with Eq.~\ref{equ:meansize}, we see that $\overline{a}\,=\,\langle\,a\,\rangle$ only if $n_{\rm d}^{\rm Gauss}\,=\,\rho L_0\,=\,n_{\rm d}$.
While this demonstrates that $n_{\rm d}^{\rm Gauss}$ must be related to the total local cloud particle number density for a monodisperse distribution ($n_{\rm d}$) because $L_{0}$ is used as the closure condition of the moment equations (see Eq.~11 in \citet{Helling2008b}), we did not use it to define $n_{\rm d}^{\rm Gauss}$. Hence $\overline{a}$ and $\langle\,a\,\rangle$ are not analytically the same, but for the atmospheres we considered, we find that they differ by no more than 20\% in the most extreme case (Fig.~\ref{fig:radii_comp}).

We calculated the opacity of non-monodisperse cloud particles for each atmospheric layer using the effective medium theory and the Mie theory as in Sect. \ref{sec:Cloud_form} to determine the dimensionless efficiency factors $Q_{\rm sca,ext,abs}(a)$ as a function of particle size within the distribution. Integrating gives the mean cross-sections for scattering, extinction, and absorption ($\langle\,C_{\rm sca,ext,abs}\,\rangle\,[\rm{cm^{2}}]$) as

\begin{equation}
    \langle\,C_{\rm sca,ext,abs}\,\rangle\,=\,\frac{\int_{a_l}^{a_u} Q_{\rm sca,ext,abs}(a)f\left(a\right)\pi a^2 da}{\int_{a_l}^{a_u} f\left(a\right)da}.
    \label{equ:optical_efficiency_integral}
\end{equation}

We numerically integrated Eq.\ref{equ:optical_efficiency_integral} by choosing upper and lower limits ($a_l$ and $a_u$) symmetric about the mean of the distribution. We find $a_{u/l} = \overline{a}\pm 5\sigma$ to be sufficient to fully capture the distribution effects. For a monodisperse distribution, that is, $f(a)\,=\,\delta(a-\langle\,a\,\rangle)$, Eq.~\ref{equ:optical_efficiency_integral} simplifies to

\begin{equation}
    C_{\rm sca,ext,abs}\,=\,Q_{\rm sca,ext,abs}(a)\pi a^2.
    \label{equ:cross_sec}
\end{equation}

The single-scattering albedo for the Gaussian distribution was similarly calculated using the integral forms as in Equation~\ref{equ:optical_efficiency_integral}.

\subsection{Particle shape}
\label{subsec:DHS}
Understanding the optical properties of non-spherical cloud particles involves modelling the interaction between the electromagnetic field and individual segments of the particles, as well as the interactions of the segments with each other. For highly non-spherical particles, it is important to take both terms into account to accurately calculate the optical properties \citep{Min2008}. Methods such as the discrete dipole approximation (DDA) \citep{Draine94,Purcell1973} model the optical properties of cloud particles with individual segments being represented by multiple dipoles. However, for very porous particles, DDA requires a large number of dipoles, $\sim\,10^{7}$ \citep{Min2008}, and thus is computationally slow.

We used the statistical approach first proposed in \citet{Bohren1983}, where the scattering and absorption effects of non-spherical grains are simulated by averaging over a distribution of simply shaped particles such as ellipsoids, spheroids, or hollow spheres \citep{Min2003}. This approach assumes that in averaging over a distribution of such shapes for a variety of parameters (e.g. major axes for an ellipsoid), the properties become approximately independent of the individual shapes used. Thus the results describe equally well what would be the case for a more thorough calculation of a distribution of irregularly shaped particles \citep{Min2003}. In the context of dust in protoplanetary discs, \citet{Min2003,Min2005} have assessed the benefits of various shape distributions. Ellipsoids, spheroids, and hollow spheres in the Rayleigh regime were described in \cite{Min2003}, and spheroids and hollow spheres for larger particles in \citet{Min2005}. The authors found good agreement between the three distributions considered for the Rayleigh regime \citep{Min2003}, and they further found that hollow spheres replicate laboratory experiments of irregularly shaped quartz particles well \citep{Min2005} for larger particles. We used the distribution of hollow spheres because it can be calculated using an extension of the Mie theory and thus yields very fast results \citep{Min2015}.

Hollow spheres are composed of two concentric spheres: a core and a mantle, with respective radii of $a_{\rm core}$ and $a_{\rm mant}$. The core is treated as a vacuum inclusion with a refractive index of unity and zero mass, and the refractive index of the mantle is taken to be that of the material cloud particles for each atmospheric layer. The total mass of the `original' compact or micro-porous cloud particle of radius $a$ (i.e. $\frac{4\pi}{3}a^{3}\rho_{\rm s}$) is distributed into the mantle. An individual hollow sphere is defined by the fraction of its total volume taken up by this core and can therefore be specified by the parameter

\begin{equation}
        f_{\rm hol}\,=\,\frac{a^{3}_{\rm core}}{a^{3}_{\rm mant}},
        \label{equ:fhol_def}
\end{equation}

hence the volume of the mantle is given by 

\begin{equation}
        V_{\rm mant}\,=\,\frac{4\pi}{3}(a^{3}_{\rm mant}-a^{3}_{\rm core}).
        \label{equ:mantle_vol}
\end{equation}

When we use Eqs.~\ref{equ:fhol_def} and \ref{equ:mantle_vol} with $\rho_{\rm s}\,=\,{\rm constant}$, the radius of the concentric spheres (core and mantle) can be written in terms of the original cloud particle radius $a$ and $f_{\rm hol}$

\begin{equation}
    a_{\rm core} = \frac{a f_{\rm hol}^{1/3}}{\left(1-f_{\rm hol}\right)^{1/3}},
    \label{equ:rcore}
\end{equation}
\begin{equation}
    a_{\rm mant} = \frac{a}{\left(1-f_{\rm hol}\right)^{1/3}}.
    \label{equ:rmant}
\end{equation}

For this paper $a$ is either taken to be the mean particle radius at an atmospheric layer, $\langle\,a\,\rangle$, or a particle radius derived from the cloud particle distribution. $a_{\rm mant}$ also defines the outer radius of the hollow sphere, and it follows from Eqs.~\ref{equ:fhol_def}, \ref{equ:rcore}, and \ref{equ:rmant}  that in the limit of $f_{\rm hol} \rightarrow 1,$ the radii of both spheres tend to infinity. Furthermore, the fraction of the hollow sphere taken up by the core by definition approaches 1. This results in an unphysical infinitely large particle consisting almost entirely of vacuum, with an infinitesimally thin mantle. This is computationally intractable for particles not in the Rayleigh regime \citep{Min2005}. Thus the distribution of hollow spheres is specified by the irregularity parameter ($f_{\rm hol}^{\rm max}$) (the upper limit of $f_{\rm hol}$ ), which should be set to a value sufficiently close to 1 so that the calculated optical properties converge whilst still remaining computationally feasible. We find $f_{\rm hol}^{\rm max} = 0.85$ to be sufficient. We used the same approach as in \citet{Min2005} (who used $f_{\rm hol}^{\rm max} = 0.98$). We averaged over a distribution function $n(f_{\rm hol})$ with equal weighting between $f_{\rm hol} = 0$ and $f_{\rm hol} = f_{\rm hol}^{\rm max}$

\begin{equation}
        n(f_{\rm hol}) = 
                \begin{cases}
                        1/f_{\rm hol}^{\rm max}, & 0 \leq f_{\rm hol} < f_{\rm hol}^{\rm max} \\
                        0, & f_{\rm hol} \geq f_{\rm hol}^{\rm max}
                \end{cases}.
\end{equation}

\section{Mineral snowflakes: effects of micro-porosity}
\label{sec:Por_effects}
We study how micro-porosity affects the cloud structure and properties of cloud particles in exoplanet and brown dwarf atmospheres. In Section~\ref{subsec:Por_material_effects} we discuss our results for the case of a warm gas giant atmosphere with $T_{\rm eff}\,=\,1800\,{\rm K}$ and $\log(\varg)\,=\,3.0$. Section~\ref{subsec:Por_albedo_effects} discusses the spectral effects of micro-porous cloud particles for this atmosphere. Finally, Section~\ref{subsec:Temp_grav} examines micro-porosity in the context of a grid of atmospheres across a range of effective temperatures and surface gravities. The 1D {\sc Drift-Phoenix} ($T_{\rm gas}$-$p_{\rm gas}$) profiles monotonically increase from $750-2500\,{\rm K}$ (Fig.~\ref{fig:DriftPhoenix}), much like nightside profiles for ultra-hot Jupiters such as WASP-18b \citep{Helling2019a} and HAT-P-7b \citep{Helling2019b}.

\begin{figure*}
    \includegraphics[page=6,width=0.5\textwidth]{Figures/Dust_Porosity/Warm_Gas_Giant_LRGRTXT.pdf}
    \includegraphics[page=2,width=0.5\textwidth]{Figures/Dust_Porosity/Warm_Gas_Giant_LRGRTXT.pdf}
    \includegraphics[page=3,width=0.5\textwidth]{Figures/Dust_Porosity/Warm_Gas_Giant_LRGRTXT.pdf}
    \includegraphics[page=4,width=0.5\textwidth]{Figures/Dust_Porosity/Warm_Gas_Giant_LRGRTXT.pdf}
    \includegraphics[page=5,width=0.5\textwidth]{Figures/Dust_Porosity/Warm_Gas_Giant_LRGRTXT.pdf}
    \includegraphics[width=0.5\textwidth]{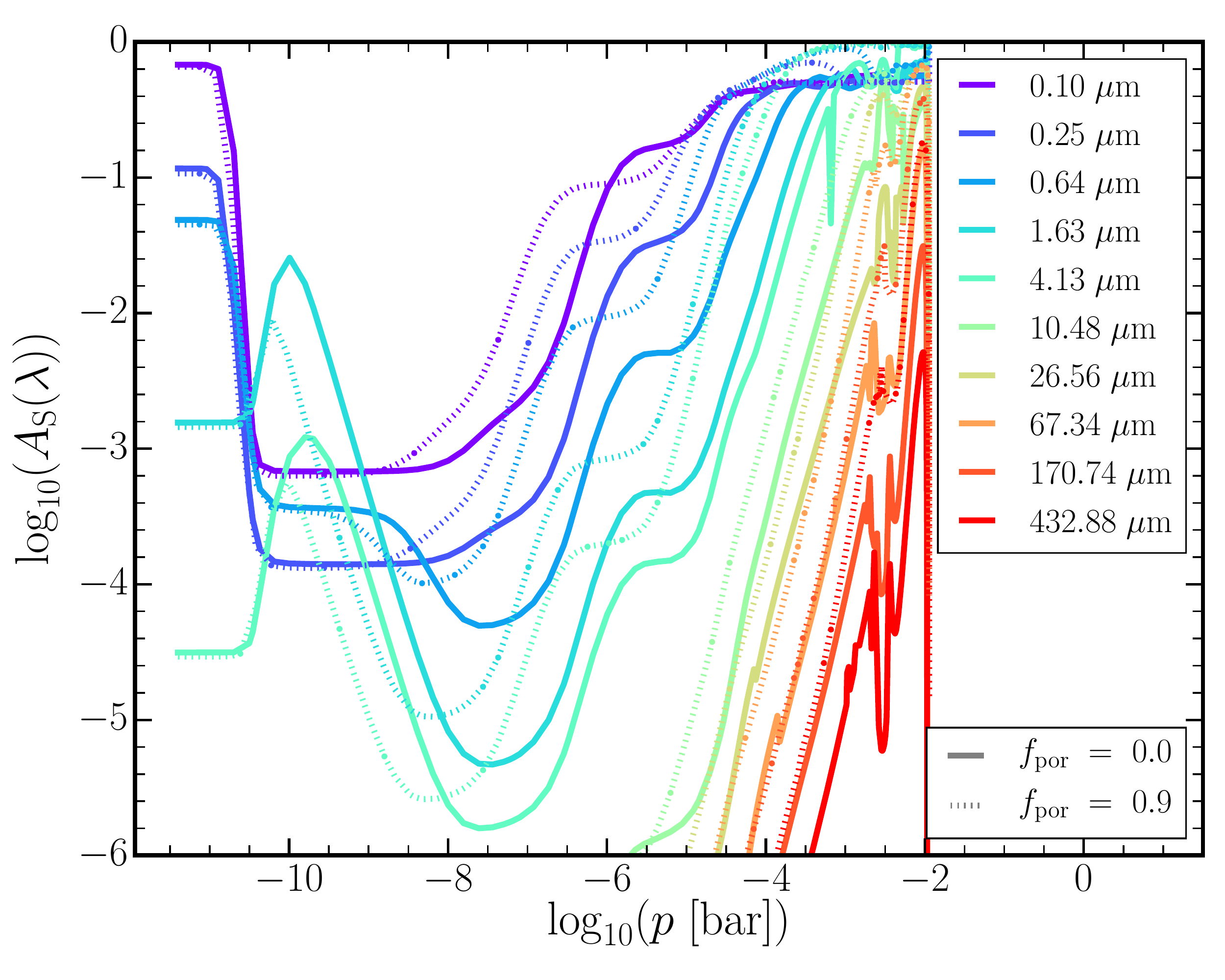}
    \caption{Material and optical properties of cloud particles in an atmosphere representative of a warm gas giant, $T_{\rm eff}\,=\,1800\,K,\, \log(\varg)\,=\,3.0$, for a cloud particle micro-porosity $f_{\rm por}\,=\,0.0,\,0.5,\,0.9$  (the bottom right panel  shows only $f_{\rm por}\,=\,0.0,\,0.9$ for clarity). \textbf{Top left:} Nucleation rates for individual species $J_{\rm i}\,[{\rm \,s^{-1}}]$ i~=~\ce{TiO2}[s] (blue), \ce{SiO}[s] (brown), and the total $J_{*} = \Sigma_{\rm i}J_{\rm i}$ (red). \textbf{Top right:} Number density $n_{\rm d}\,[\rm{cm^{-3}}]$ of cloud particles. \textbf{Middle left:} Mean cloud particle size $\langle\,a\,\rangle\,[\rm{\mu m}]$. \textbf{Middle right:} Mean drift velocity $\langle\,v_{\rm dr}\,\rangle \,[\rm{cm\,s^{-1}}]$ for cloud particles of size $\langle\,a\,\rangle$. \textbf{Bottom left:} Ratio of the cloud particle mass density and gas mass density ($\rho_{\rm{d}}/ \rho$) scaled by a factor of $10^{-3}$. \textbf{Bottom right:} Single-scattering albedo as defined in Eq.~\ref{equ:ssa} for wavelengths $\lambda = 0.1,\,0.25,\,0.64,\,1.63,\,4.13,\,10.48,\,26.56,\,67.34,\,170.74,\text{and }\,432.88\,\rm{\mu m}$ (colours as shown in the legend).}
    \label{fig:multi_porous}
\end{figure*}

\subsection{Micro-porosity and amplified bulk growth}
\label{subsec:Por_material_effects}
For the warm gas giant atmosphere ($T_{\rm eff} = 1800\,{\rm K},\,\log(\varg)\,=\,3.0$) we considered three values for the micro-porosity ($f_{\rm por}\,=\,0.0,0.5,0.9$). Increasing the micro-porosity of the cloud particles leads to generally larger cloud particles (Fig.~\ref{fig:multi_porous}, middle left). In the upper atmosphere this is due to the increased nucleation monomer size: at greater pressures, the mean cloud particle size increases at $\sim\,10^{-8}\,{\rm bar}$ for $f_{\rm por}\,=\,0.9$ as opposed to $\sim\,10^{-7}\,{\rm bar}$ for compact particles ($f_{\rm por}\,=\,0.0$). This is due to the larger surface area of the cloud particles, which increases the altitude at which bulk growth begins. Figure~\ref{fig:mat_comp} shows the material composition of cloud particles, the transition between cloud particles composed entirely of the nucleation species \ce{SiO}[s] to cloud particles with a significant fraction of the bulk growth material \ce{MgO}[s] shifts to higher in the atmosphere. This transition shifts from $\sim\,10^{-7}\,{\rm bar}$ to $\sim\,10^{-8}\,{\rm bar}$ between the compact (Fig.~\ref{fig:mat_comp}, left) and highly porous (Fig.~\ref{fig:mat_comp}, right) cases. When we simplify this by assuming constant material composition, the drift velocity of a micro-porous particle can be expressed in terms of the compact drift velocity and the micro-porosity fraction (from Eqs.~\ref{equ:driftv} and \ref{equ:effden}),

\begin{equation}
        v_{\rm dr}^{\rm por} = \left(1-f_{\rm por}\right)v_{\rm dr}^{\rm comp}.
        \label{equ:por_drift}
\end{equation}

Lower drift velocities are therefore expected for higher micro-porosity, as shown in the middle right panel of Fig.~\ref{fig:multi_porous}. This furthermore allows the cloud particles to remain longer in an atmospheric layer and thus to experience more bulk growth. It therefore causes greater particle sizes. Conversely, it does not lead to a significant increase in the ratio of cloud particle mass to gas mass $\rho_{\rm d}/\rho$ (Fig.~\ref{fig:multi_porous}, bottom right) because it is balanced by a reduced number density: with increasing micro-porosity there are fewer large cloud particles in any given atmospheric layer.

The micro-porosity also affects the number of seed particles that form, that is, the nucleation rate (Fig.~\ref{fig:multi_porous}, top left). The nucleation rates of \ce{TiO2}[s] and \ce{SiO}[s] decrease with increased cloud particle micro-porosity. However, the micro-porosity must reach 90\% for the nucleation rate to decrease significantly, and below 50\% micro-porosity, the effect remains within an order of magnitude for all pressures. Micro-porosity does not affect the rate at which cloud particles form in the low-pressure atmosphere. The point at which the nucleation rate deviates from the compact particle rate ($f_{\rm por}\,=\,0.0$) is dependent on the cloud particle micro-porosity and occurs at higher pressures for increased micro-porosity. This is due to competitive bulk growth rates for the nucleation elements. Because bulk growth needs substantially less supersaturation of the gas phase than nucleation, the bulk growth rate quickly exceeds the nucleation rate when the material becomes thermodynamically stable. This depletes the gas phase of Si-, Ti-, and \ce{O} -bearing molecules and thus limits the nucleation rates. A higher particle micro-porosity increases the area that is available for gas-surface reactions and thus improves the bulk growth rate at all layers, which means that it becomes stronger than nucleation at higher altitudes than in the compact case. The reduction in nucleation rate leads to the reduced peak number density of cloud particles in the atmosphere (Fig.~\ref{fig:multi_porous}, top right). 

\subsection{Increased albedo as a result of the micro-porosity of cloud particles}
\label{subsec:Por_albedo_effects}
\begin{figure}
    \includegraphics[width=0.5\textwidth,page=1]{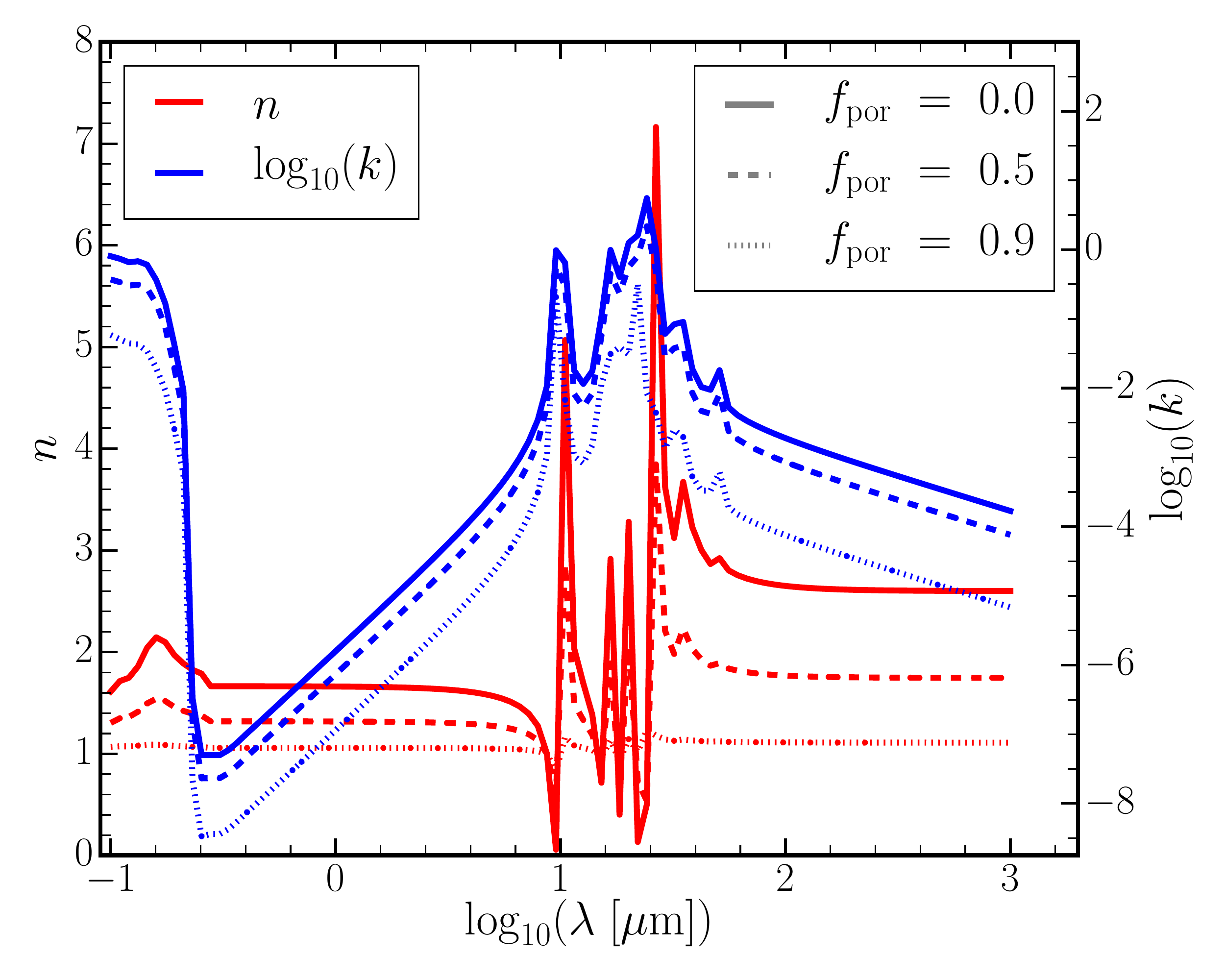}
    \caption{Real ($n$) and imaginary ($k$) refractive indices calculated based on the effective medium theory (red and blue lines, respectively) for cloud particles consisting of crystalline \ce{Mg2SiO4}[s] and vacuum, for micro-porosity fractions $f_{\rm  por} = 0.0,\,0.5,\,0.9$.}
    \label{fig:Porousfosterite}
\end{figure}

In order to discuss the optical effects of micro-porosity on cloud particles, it is illuminating to first briefly consider homogeneous cloud particles that are composed of only one material and vacuum through the effective medium theory. Figure~\ref{fig:Porousfosterite} shows that as $f_{\rm por}$ increases, the refractive index tends towards vacuum values, with the real refractive index becoming more uniform (diminishes spectral features) and tending to 1, whilst the imaginary component also decreases uniformly, although it maintains its shape, and tends towards zero.

For the warm gas giant atmosphere, the shape of the single-scattering albedo ($A_{\rm S}(\lambda)$, Eq.~\ref{equ:ssa}) `spectrum' (Fig.~\ref{fig:multi_porous}, bottom right) is defined by the material composition of the cloud particles. The sudden drop in $A_{\rm S}(\lambda)$  at $10^{-10}\,{\rm bar}$ for wavelengths $\lambda = 0.1,\,0.25,\text{and}\,0.64\,{\rm \mu m}$ is associated with the transition between the dominant nucleation species that changes from cloud particles consisting entirely of \ce{TiO2}[s] seeds to particles that largely comprise \ce{SiO}[s], which aligns with the dominant nucleation rates (Fig.~\ref{fig:multi_porous}, top left). The exact values of the albedo in this region and the transition is dependent on the extrapolation used for \ce{TiO2}[s] because the refractive index data do not cover this wavelength range (see Fig.~\ref{fig:appendix_refindex2}). Further work on obtaining the refractive indices for materials such as \ce{TiO2}[s] over a broader range of wavelengths would greatly benefit many theoretical studies. The increase in albedo for all wavelengths between $10^{-8}$ and $10^{-7}\,{\rm bar}$ is associated with a change to heterogeneous cloud particles of mixed composition, with a majority iron and magnesium silicate composition. Figure~\ref{fig:mat_comp} shows detailed material composition changes across the atmosphere for the compact and highly porous cases. Below $10^{-7}\,{\rm bar,}$ the albedo for wavelengths smaller than $10.48\,{\rm \mu m}$ shows a general trend corresponding to the mean grain size. Most notably, this explains the shift of the flat part of the spectrum from around $10^{-6}\,{\rm bar}$ for the compact case to $10^{-7}\,{\rm bar}$. In this flattened region the effects of micro-porosity are directly visible on the albedo, with increased values for all wavelengths except $0.1\,{\rm \mu m}$. For the highly micro-porous particles the effective refractive index tends towards that of vacuum and the extinction efficiency of the particles ($Q_{\rm{ext}}$) is therefore reduced, increasing the albedo. Whilst the extinction efficiency of the cloud particles is reduced, the extinction cross-section Eq.~\ref{equ:cross_sec} of micro-porous cloud particles can still increase because of their larger size (see Fig.~\ref{fig:holo_albedo_lambda}).

\begin{figure*}
    \includegraphics[page=2,width=0.5\textwidth]{Figures/TempGrav/Temp_Exploration_LRGRTXT.pdf}
    \includegraphics[page=1,width=0.5\textwidth]{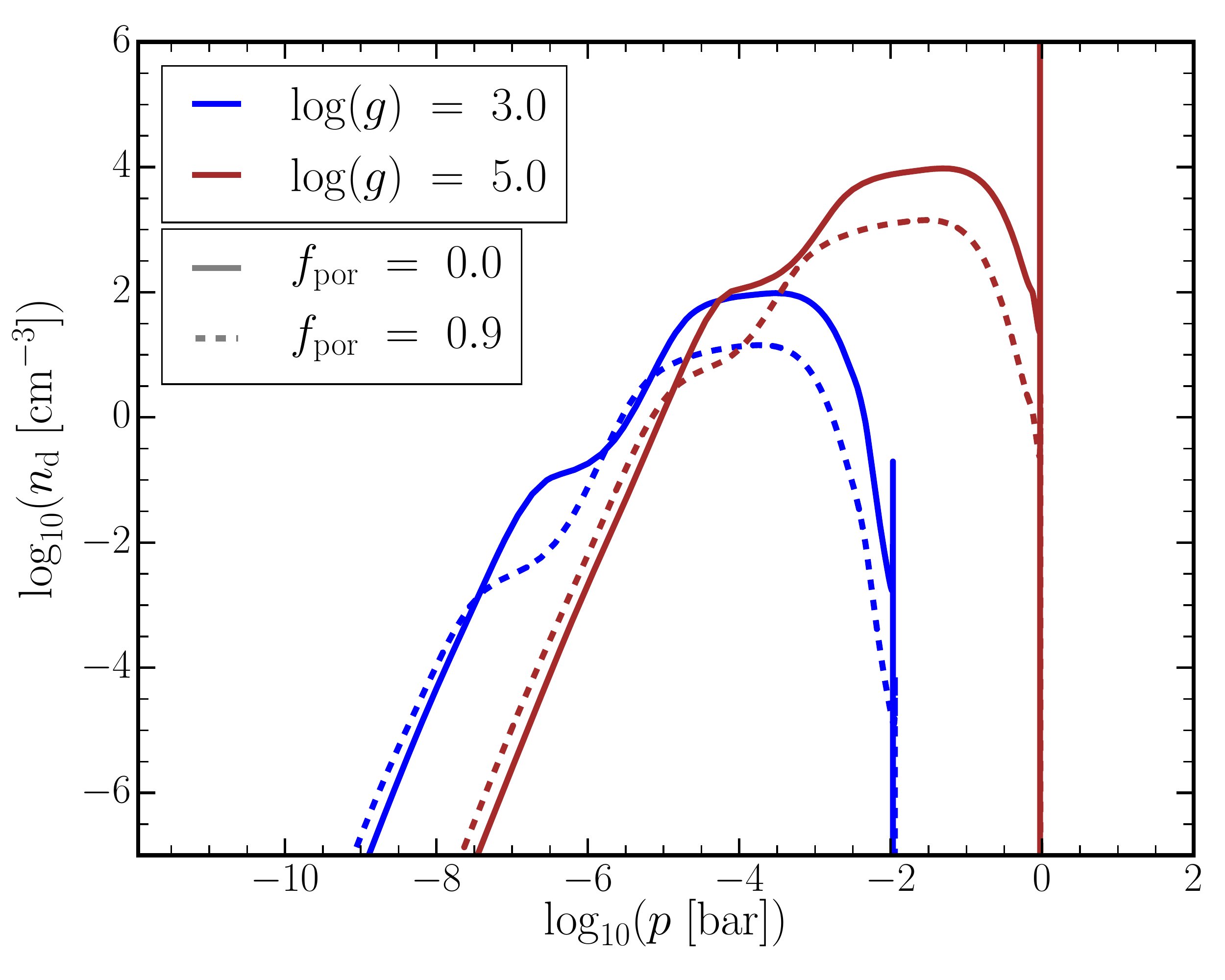}
    \includegraphics[page=3,width=0.5\textwidth]{Figures/TempGrav/Temp_Exploration_LRGRTXT.pdf}
    \includegraphics[page=2,width=0.5\textwidth]{Figures/TempGrav/Grav_Exploration_LRGRTXT.pdf}
    \includegraphics[page=5,width=0.5\textwidth]{Figures/TempGrav/Temp_Exploration_LRGRTXT.pdf}
    \includegraphics[page=4,width=0.5\textwidth]{Figures/TempGrav/Grav_Exploration_LRGRTXT.pdf}
    \caption{Material properties of cloud particles for micro-porosities of $f_{\rm por} = 0.0$ (solid) and $f_{\rm por} = 0.9$ (dashed) across a grid of effective temperatures and surface gravities.
    \textbf{Left column:} Surface gravity $\log(\varg) = 3.0$, representative of gas giant exoplanets, for $T_{\rm eff}\,=\,1200,\,1800,\text{ and}\,2400\,{\rm K}$ (red, blue, and green, respectively).
    \textbf{Right column:} Surface gravities $\log(\varg)\,=\,3.0$ (blue) and $5.0$ (brown), representative of brown dwarfs and young gas giants for $T_{\rm eff}\,=\,1800\,{\rm K}$.
    The blue lines in both columns are identical.
    \textbf{Top row:} Number density of cloud particles $n_{\rm d}\,[\rm{cm^{-3}}].$
    \textbf{Middle row:} Mean cloud particle size $\langle\,a\,\rangle\,[\rm{\mu m}]$.
    \textbf{Bottom row:} Ratio of the cloud particle mass density and gas mass density $\rho_{\rm{d}}/\rho$ scaled by a factor of $10^{-3}$.}
    \label{fig:temp_grav_material}
\end{figure*}

\begin{figure}
    \includegraphics[page=1,width=0.5\textwidth]{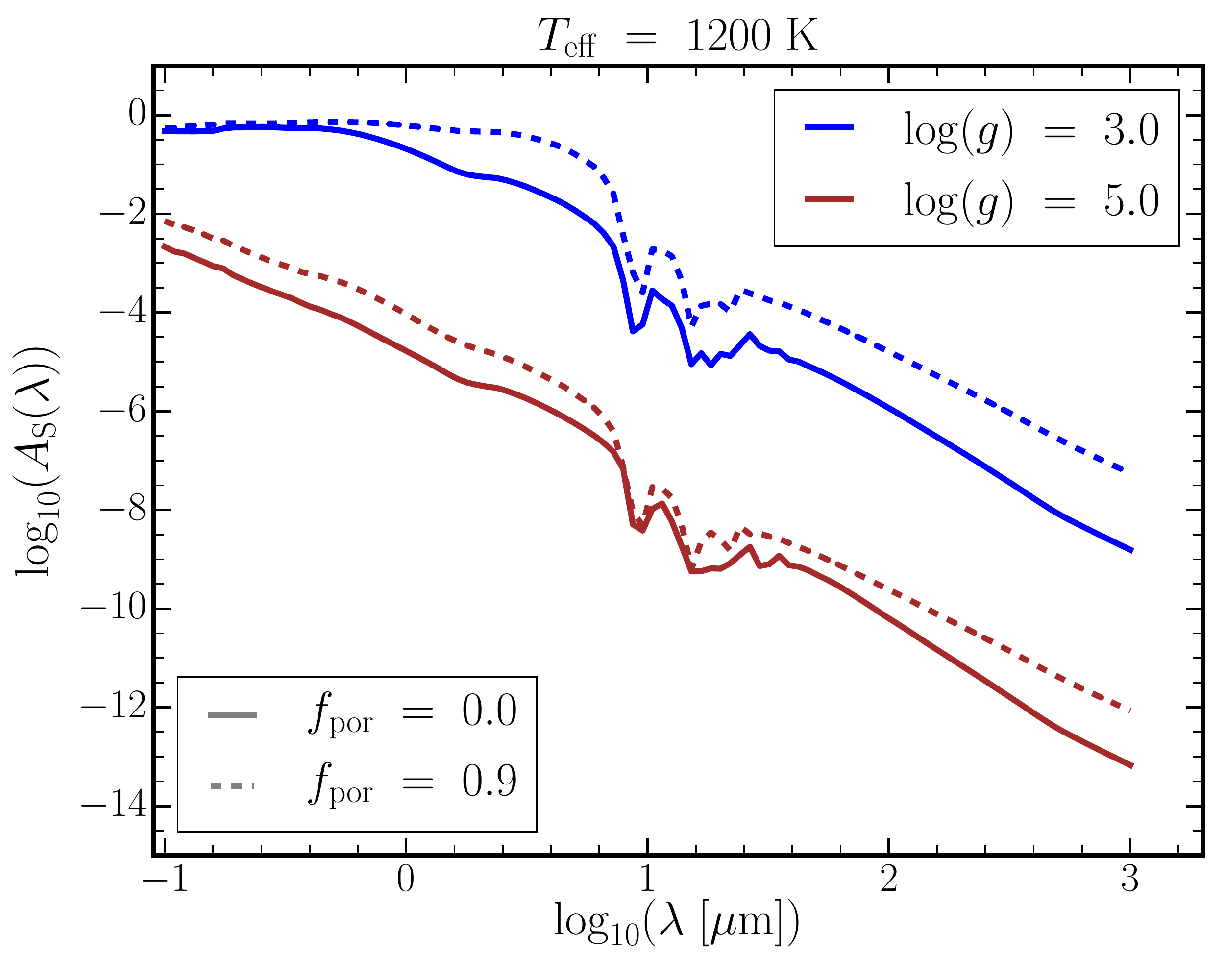}
    \includegraphics[page=3,width=0.5\textwidth]{Figures/TempGrav/Albedo_Exploration_LRGRTXT.pdf}
    \includegraphics[page=5,width=0.5\textwidth]{Figures/TempGrav/Albedo_Exploration_LRGRTXT.pdf}
    \caption{Single-scattering albedo $A_{\rm S}(\lambda)$ as a function of wavelength from $0.1\,\rm{\mu m}$ to $1000\,\rm{\mu m}$ for $f_{\rm por}\,=\,0.0$ (solid) and 0.9 (dashed) at pressure level of $1\,{\rm mbar}$. Surface gravities of $\log(\varg)\,=\,3.0$ (blue) and $5.0$ (brown) are shown. \textbf{Top}, \textbf{middle,} and \textbf{bottom} panels are for $T_{\rm eff}\,=\,1200,\,1800,\text{and}\,2400\,{\rm K,}$ respectively.}
    \label{fig:temp_grav_albedo}
\vspace*{-5mm}
\end{figure}

\subsection{Effect of micro-porosity for different $T_{\rm eff}$ and $\log(\varg)$}
\label{subsec:Temp_grav}
The trends in $n_{\rm d},\,\langle\,a\,\rangle,\,\rm{and}\,\rho/\rho_{\rm d}$ as previously observed for the ${\rm T}_{\rm eff}\,=\,1800\,{\rm K}$ gas giant atmosphere qualitatively hold for a wider set of global atmospheric parameters (Fig.~\ref{fig:temp_grav_material}). For all atmospheres we investigated, the increased surface area of the micro-porous cloud particles leads to growth that occurs higher in the atmosphere and consequently to a reduced peak in number density $n_{\rm d}$ of cloud particles. In the upper atmosphere of each profile, the number density of micro-porous particles initially marginally exceeds the compact case, but when growth becomes efficient, it dominates nucleation for the necessary nucleation elements and thus dramatically reduces the nucleation rates and lowers the peak cloud particle number density. The ratio of cloud mass to gas mass, $\rho_{\rm d}/\rho$, peaks deeper in the atmosphere of cooler planets because the temperatures sufficient to evaporate the condensation species occur at greater pressure for these atmospheres. Because bulk growth rates and drift velocities are lower, atmospheres that are typically inefficient at sequestering mass into cloud particles have a significantly higher ratio of cloud particle mass to gas mass, such as the $T_{\rm eff}\,=\,2400\,{\rm K}$, $\log(\varg)\,=\,3.0$ profile. The bottom left plot of Figure~\ref{fig:temp_grav_material} shows that in the compact case, the ratio of peak cloud mass to gas mass of this profile is about $0.5 \times 10^{-3}$, whereas in the highly micro-porous case the peak value is closer to $1 \times 10^{-3}$. This is an increase of a factor of 2. The increase for the $T_{\rm eff}\,=\,1200\,{\rm K}$ profile, which is much more efficient at cloud particle formation in the compact case, is only minor. Higher surface gravity furthermore causes cloud formation to occur deeper in the atmosphere (Fig.~\ref{fig:temp_grav_material}, right) because the gas pressure is higher, but the increase in $\rho_{\rm d}/\rho$ remains roughly consistent for both cases. The average particle size similarly follows the change in cloud formation when $\log(\varg)\,=\,3.0$ and $5.0$ are compared. Similarly, when we varied the temperature for $\log(\varg)\,=\,3.0$ profiles, the $T_{\rm eff}\,=\,1600\,{\rm K}$ model and the $T_{\rm eff}\,=\,2400\,{\rm K}$ model increase the average cloud particle size by about the same amount as for the $T_{\rm eff}\,=\,1800\,{\rm K}$. However, for $T_{\rm eff}\,=\,2400\,{\rm K,}$ the average particle size remains close to the same size as for the $T_{\rm eff}\,=\,1800\,{\rm K}$ case for pressures greater than $10^{-6}\,{\rm bar}$  because the two are similar even for the compact case.

Figure~\ref{fig:temp_grav_albedo} shows the optical effects of micro-porous particles for our grid of atmospheres at the $1\,{\rm mbar}$ pressure level. this region of the atmosphere is typically probed by transmission observations for gas giant planets. Generally, the single-scattering albedo is enhanced for all profiles and across all wavelengths we considered, with the albedo for a brown dwarf profile always lower than that of a similar temperature gas giant profile. For wavelengths shorter than $10\,\rm{\mu m}$ in the $T_{\rm eff}\,=\,2400\,{\rm K}$ brown dwarf model atmosphere, there is a peak increase of two orders of magnitude in albedo around $1\,{\rm \mu m}$. For all gas giant profiles, the albedo for wavelengths around $10\,\rm{\mu m}$ becomes flatter. For longer wavelengths ($>\,30\,{\rm \mu m}$), the micro-porous particles show a trend of increased albedo over the compact case, with an increase of roughly two orders of magnitude for the hottest profiles at wavelengths of a few hundred microns. Lastly, we note an increased prominence of silicate features (which make up the bulk of the volume of cloud particles at these pressures) for all profiles in the micro-porous case. 

Across the range of models we find that the increases in various material properties of the cloud particles across the atmosphere remain the same for various temperatures and surface gravities. This was expected because we chose to model micro-porosity by a constant factor. A parameter dependent on local temperature and gas density would capture effects such as different evaporation rates of materials, leading to porous inclusions in the cloud particle. These effects would increase the micro-porosity deeper in the atmosphere.

\section{Cloud particle size distribution, high-altitude cloud material, and optical properties}
\label{sec:Sizedist_effects}
The effects of a height-dependent cloud particle size distribution are shown in Fig.~\ref{fig:Sizedist} for the $T_{\rm eff}\,=\,1800\,{\rm K},\,\log(\varg)\,=\,3.0$ atmosphere. The functional form is a Gaussian distribution, derived from the solution of our kinetic cloud formation model as described in Section~\ref{subsec:Sizedist}. For simplicity, we considered only compact cloud particles in this section. The coupled effects of micro-porosity and non-monodisperse cloud particles are discussed in Section~\ref{sec:Coupled_effects}.

\subsection{Wide particle size distributions due to competitive growth and nucleation}
\label{subsec:Sizedist_material_effects}
\begin{figure*}
    \includegraphics[page=3,width=0.5\textwidth]{Figures/Dust_Size_Distributions/Sizedist_grid_LRGRTXT.pdf}
    \includegraphics[page=1,width=0.5\textwidth]{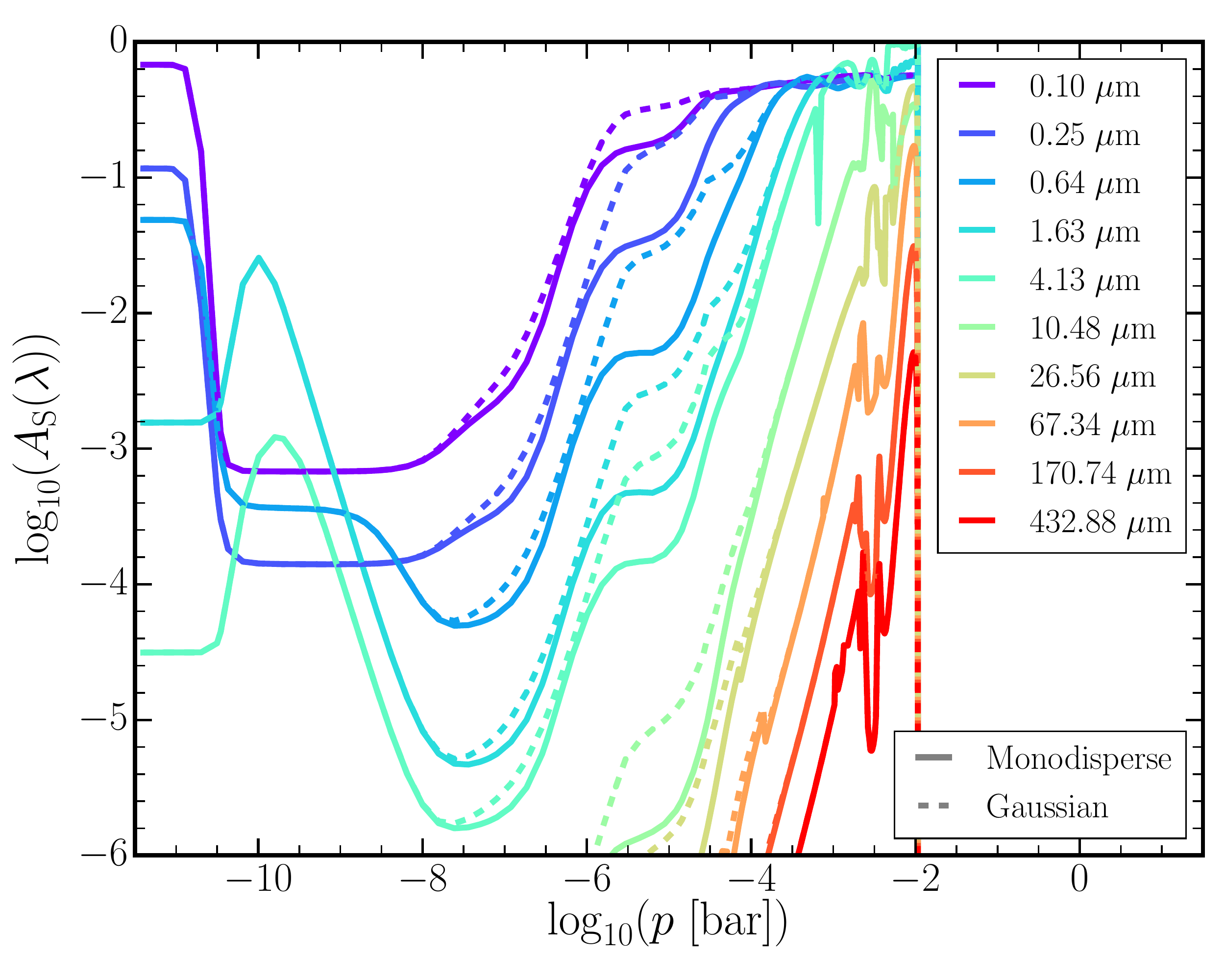}
    \includegraphics[page=2,width=0.5\textwidth]{Figures/Dust_Size_Distributions/Sizedist_grid_LRGRTXT.pdf}
    \includegraphics[page=4,width=0.5\textwidth]{Figures/Dust_Size_Distributions/Sizedist_grid_LRGRTXT.pdf}
    \caption{Gaussian cloud particle size distribution material and optical effects (atmosphere of $T_{\rm eff}\,=\,1800\,{\rm K},\,\log(\varg)\,=\,3.0$). \textbf{Top left:} Gaussian distribution mean cloud particle size $\overline{a}\,\rm{\mu m}$ (solid black). Green contours show particle sizes $1\,\sigma,\,3\,\sigma,\ {\rm and}\ 5\,\sigma$ from the mean size (left axis). \textbf{Top left:} Cloud particle number density $n_{d}\,[\rm{cm^{-3}}]$ (dashed blue line, right axis). Below this we reproduce material property plots to facilitate comparison. These are identical to those shown in Fig.~\ref{fig:multi_porous}. \textbf{Bottom left:} Nucleation rates (left axis) for individual species $J_{\rm i}\,[{\rm cm^{3}\,s^{-1}}]$ i = \ce{TiO2}[s] (blue), \ce{SiO}[s] (brown), and the total $J_{*}\,=\,\Sigma_{\rm i}J_{\rm i}$ (red). \textbf{Bottom left:} Ratio of the cloud particle mass density and gas mass density $\rho_{\rm{d}}/ \rho$ scaled by a factor of $10^{-3}$ (black dashed line, right axis). \textbf{Top right:} Single-scattering albedo ($A_{\rm S}(\lambda)$) for monodisperse cloud particles $\langle\,a\,\rangle\,[\rm{\mu m}]$ (solid) and Gaussian size distribution (dotted) at wavelengths $\lambda = 0.1,\,0.25,\,0.64,\,1.63,\,4.13,\,10.48,\,26.56,\,67.34,\,170.74,\text{and}\,432.88\,\rm{\mu m}$. \textbf{Bottom right:} Ratio of the single-scattering albedo for a Gaussian and monodisperse size distribution ($A_{\rm S}^{\rm Gauss}/A_{\rm S}^{\rm Mono}$), i.e. a value of $>\,1$ indicates an increase in the albedo when a particle size distribution is included. The wavelengths are the same as for top right plot.}
    \label{fig:Sizedist}
\end{figure*}

The top left panel of Figure~\ref{fig:Sizedist} shows the deviation of the Gaussian distributed particle sizes from a simple delta-function-like monodisperse ($a\,=\,\langle\,a\,\rangle$) distribution. For the upper atmosphere ($<\,10^{-8}\,{\rm bar}$), only condensation seeds nucleate, and thus the distribution is almost a delta function around $\overline{a}$. Particles do not undergo bulk grow in this regime as the growth timescale is much longer than the gravitational settling timescale ($\tau_{\rm{gr}} >> \tau_{sink}$). The cloud particles therefore rapidly fall before significant growth can occur. The mean of the size distribution at each level in the atmosphere therefore remains at a constant value, the size of the nucleating seeds, down to $10^{-8}\,{\rm bar,}$ as was the case for monodisperse cloud particles. Furthermore, in this region we see the familiar effects of changing material composition, which remains unaffected by local particle sizes, as discussed in Section~\ref{subsec:Por_albedo_effects}.

Around $10^{-8}\,{\rm bar}$, $\tau_{\rm{gr}} = \tau_{sink}$, therefore cloud particles begin to grow through the condensation of thermally stable materials. In this region the mean cloud particle size begins to increase, but the nucleation rate remains high (Fig.~\ref{fig:Sizedist}, bottom left); the rate of \ce{SiO}[s] nucleation does not peak until approximately $10^{-7}\,{\rm bar}$ (brown line). Thus the inferred local variance of the Gaussian distribution from the moments increases significantly to account for both the small nucleation seeds and for particles beginning to undergo substantial bulk growth in the same atmospheric layer. A similar broadening is found at $10^{-5}\,{\rm bar,}$ where the \ce{TiO2}[s] nucleation rate peaks (blue line). In between these two peaks, the nucleation rate briefly drops. This leads to a plateau in the particle number density (blue dashed line Fig.~\ref{fig:Sizedist}, top left). At this same point, the distribution narrows because the cloud particles grow rapidly. Any cloud particle seeds falling from higher in the atmosphere in this regime ($>\,10^{-8}\,{\rm bar}$) rapidly grow, and thus a significant population of small particles is only supported by high nucleation rates. Below $10^{-4}\,{\rm bar,}$ the total nucleation rate drops rapidly and the Gaussian distribution converges towards the mean cloud particle size $\langle\,a\,\rangle$. 

In regions where the local cloud particle size distribution is represented by a wide Gaussian distribution, the symmetrical nature of the Gaussian distribution can lead to unphysical inferred particle sizes that extend below the minimum particle size and indeed even below zero in attempting to represent the large particles produced by bulk growth. In the atmosphere we studied, this is the case even for particles within $3\sigma$ of the mean for the region between $10^{-6}$ and $10^{-4}\,{\rm bar}$. Cloud particle sizes below the minimum particle size are not included in any of the material property calculations because these rely on the moments, and using the distribution to calculate the optical properties therefore produces a discrepancy between the methods. Cloud particles in the size distribution below zero size are excluded, and a further small discrepancy is therefore produced in the total number density of cloud particles. It is also unclear to what extent the upper bounds of the distribution are increased in attempting to compensate for the small nucleation seeds over the true values for the larger cloud particles that have undergone bulk growth. Approaches to cloud particle and haze formation through the binning method have found evidence for multi-modal distributions, but these models rely on assumptions that simplify the formation of seed particles, material compositions, and growth processes \citep{Powell2018,Kawashima2018}.

\subsection{Increased albedo due to non-monodisperse cloud particle size distribution}
\label{subsec:Sizedist_albedo_effects}
At pressures where the distribution is narrow, the integrated albedo over the size distribution is the same as for the monodispere case. For the broad distribution between $10^{-8}$ and $10^{-4}\,{\rm bar}$, the single-scattering albedo of the Gaussian distribution ($A^{\rm{Gauss}}_{\rm S}(\lambda)$) is always higher than that of the monodisperse case ($A^{\rm{Mono}}_{\rm S}(\lambda)$), see the right two plots of Fig.~\ref{fig:Sizedist}. The increase for the initial broadening around $10^{-7}\,{\rm bar}$ is wavelength independent and peaks at $A^{\rm{Gauss}}_{\rm S}\,=\,1.5 A^{\rm{Mono}}_{\rm S}$ (Fig.~\ref{fig:Sizedist}, bottom right). The second peak in the increase of the albedo from the monodispserse case occurs at about $10^{-5}\,{\rm bar,}$ and this time is wavelength dependent with a significantly lower increase of only a factor of 2 for $\lambda\,=\,0.1\,{\rm \mu m,}$ as opposed to increases by a factor greater than 4 for all longer wavelengths. This is because in this region $\overline{a}\,=\,10^{-2}\,{\rm [\mu m],}$ and the fraction of cloud particles in the Rayleigh scattering regime ($Q_{\rm{sca}}\,\propto\,\lambda^{-4}$) varies for each wavelength for a broad size distribution. Although an increase of half an order of magnitude in albedo seems substantial when the total extinction of the atmosphere is calculated, the peak extinction is dominated by regions with high mass fractions of cloud particles ($\rho_{\rm d}/\rho$). The mass fraction of cloud particles peaks around the millibar level in the atmosphere, which does not significantly overlap with regions of high nucleation rates (Fig.~\ref{fig:Sizedist}, bottom left) where the local size distribution is broad. The size distribution therefore has little effect on properties such as optical depth (Fig.~\ref{fig:Opt_depth}). \citet{Wakeford2015} also found that for a log-normal distribution, the cumulative effect of transmission spectra of clouds is dominated by larger particles in the distribution. An asymmetric distribution, such as the log-normal distribution, has a greater number of large particles than the Gaussian distribution considered here, and may produce greater deviations from the monodisperse case. It has been found recently to have an effect on retrieved cloud particle size \citep{Benneke2019a}. We note, however, that none of the size distributions used in retrieval approaches are based on a consistent cloud model, and therefore no conclusion can be drawn about the shape of the particle size distribution from such retrieval approaches.

\section{Optical effects of non-spherical particles}
\label{sec:DHS_optical_effects}
\begin{figure*}
    \includegraphics[page=1,width=0.5\textwidth]{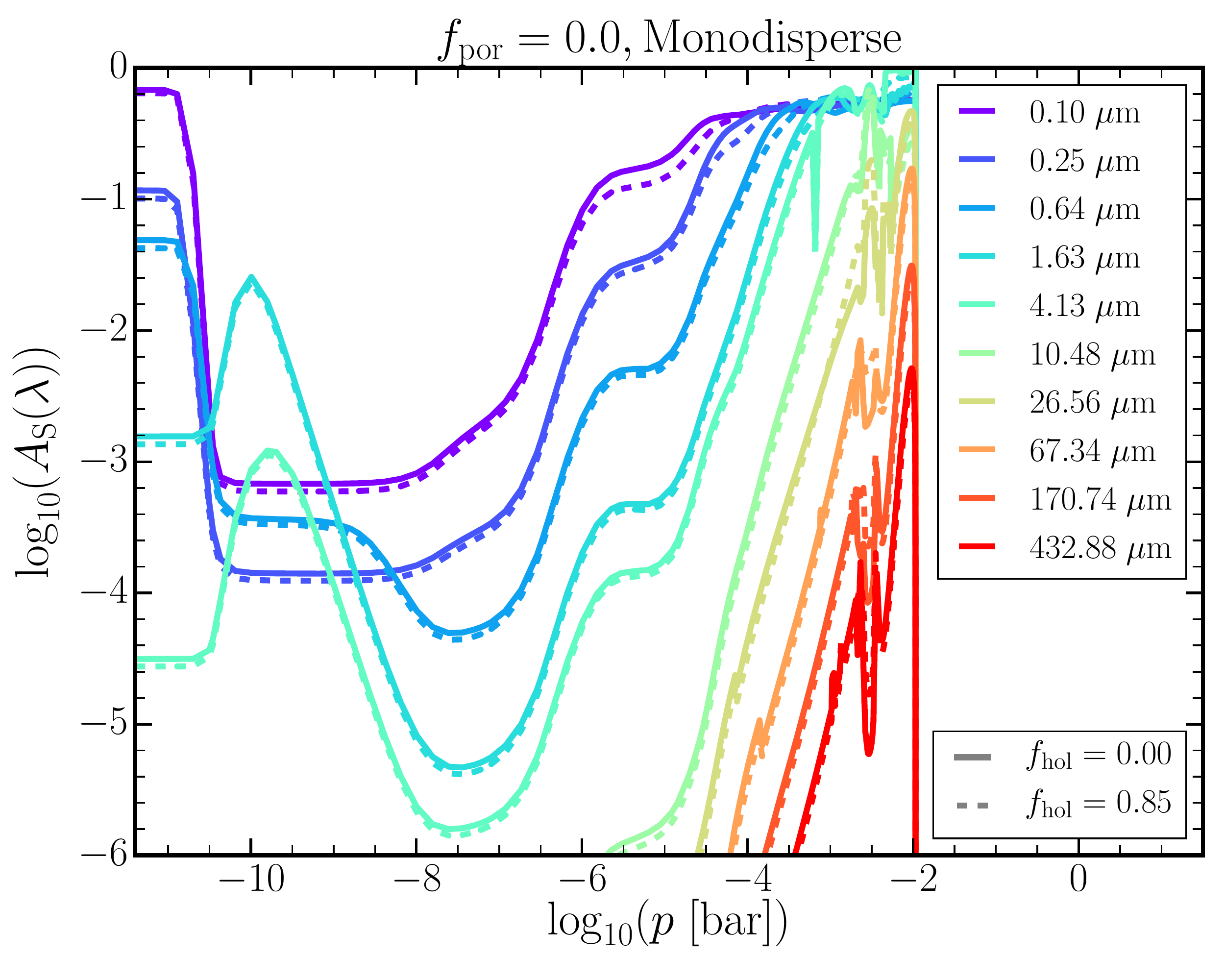}
    \includegraphics[page=2,width=0.5\textwidth]{Figures/Holospheres/Albedo_HOLO_Exploration_LRGRTXT.pdf}
    \includegraphics[page=3,width=0.5\textwidth]{Figures/Holospheres/Albedo_HOLO_Exploration_LRGRTXT.pdf}
    \includegraphics[page=4,width=0.5\textwidth]{Figures/Holospheres/Albedo_HOLO_Exploration_LRGRTXT.pdf}
    \caption{Single-scattering albedo ($A_{\rm S}(\lambda)$) with (solid) and without (dashed) a distribution of hollow spheres for the atmosphere $T_{\rm eff}\,=\,1800\,{\rm K},\,\log(\varg)\,=\,3.0$ at wavelengths $\lambda = 0.1,0.25,0.64,1.63,4.13,10.48,26.56,67.34,170.74,\text{and }432.88\,\rm{\mu m}$.
    \textbf{Left column:} For monodisperse cloud particles according to mean particle size $\langle\,a\,\rangle\,[\rm{\mu m}]$.
    \textbf{Right column:} For the derived Gaussian cloud particle size distribution.
    \textbf{Top row:} For compact cloud particles ($f_{\rm por} = 0.0$). 
    \textbf{Bottom row:} For highly micro-porous ($f_{\rm por} = 0.9$) cloud particles.}
    \label{fig:holosphere_albedo}
\end{figure*}

To determine the effect of non-spherical cloud particles, we return to the $T_{\rm eff}\,=\,1800\,{\rm K},\,\log(\varg)\,=\,3.0$ atmosphere with compact monodisperse cloud particles. When a distribution of hollow spheres is used, as described in Section~\ref{subsec:DHS}, the albedo generally decreases compared to the compact monodisperse case (Fig.~\ref{fig:holosphere_albedo}, top right). This is due to the high $f_{\rm hol}$ end of the hollow sphere distribution, where the cloud particles have a large surface area (which is not conserved across the distribution of hollow spheres) and thus increase absorption \citep{Min2003}. Furthermore, Fig.~\ref{fig:Opt_depth} shows that for the compact case (red line), the inclusion of non-spherical particles leads to an increase in optical depth in the silicate features between $10-20\,\rm{\mu m}$. This result is in good agreement with \citet{Min2005}, who found that particles from the distribution with large $f_{\rm hol}$, and thus thin mantles, act similarly to a collection of smaller particles that enhance spectral resonances. The result is more complex for hollow spheres coupled with already micro-porous particles, with little to no effect on the features because the two methods (distribution of hollow spheres and effective medium theory) both attempt to capture the effects of non-compact particles. the implications of the combined effects are therefore unclear. This shows that the limited effect of hollow spheres in addition to highly micro-porous particles is due to the diluted refractive indices that such particles have.

Our results are in qualitative agreement with the result found by \citet{Min2005}, although total agreement is not possible because the model set-ups and sources of refractive index data were different. We show that the results of retrievals for particle sizes \citep{Benneke2019a} are greatly simplified with compact spheres and that this simplification can have a dramatic effect on the derived cloud properties. We note, however, that here we do not explore the additional effects of a distribution of hollow spheres on the polarisation of light in these atmospheres, which can also be significant. Because spheres are surface-minimising volumes, it is expected that the surface area increases for non-spherical particles.

\section{Coupled effects of non-sphericity, particle size distribution, and micro-porosity}
\label{sec:Coupled_effects}
After examining each of the effects in isolation, we now investigate the effect on optical properties for both compact and highly micro-porous (90\%) particles; with and without a dispersed particle size distribution to represent the particle size, derived from our cloud model results; and for spherical and non-spherical grains as represented by a distribution of hollow spheres. All results are for the warm gas giant planet used throughout this paper.

Figure~\ref{fig:holosphere_albedo} shows the single-scattering albedo and serves to highlight the relative contribution of each of the individual deviations from compact spheres. As noted in Section~\ref{subsec:Sizedist_albedo_effects}, the Gaussian distribution has the effect of maintaining a higher albedo at greater atmospheric altitude. In all cases the inclusion of non-spherical particles with a distribution of hollow spheres reduces the albedo for short wavelengths at pressures lower than $10^{-4}$. For a wavelength of $0.1\,\rm{\mu m,}$ non-spherical particles have a more pronounced effect for the highly micro-porous case around $10^{-6}\,{\rm bar}$, but the reduction in albedo persists to even higher altitude only for the compact case.

\begin{figure*}
    \includegraphics[page=2,width=0.5\textwidth]{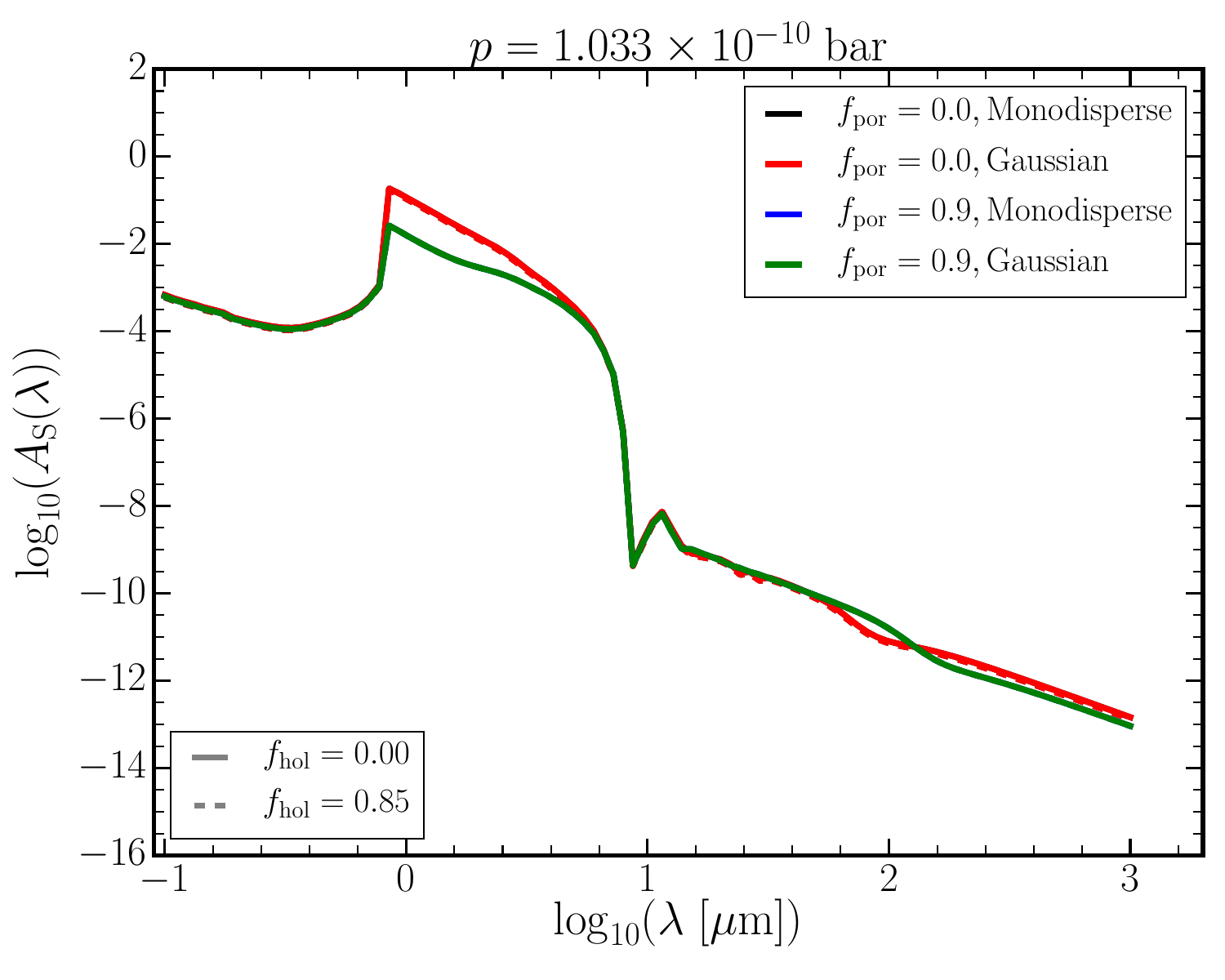}
    \includegraphics[page=2,width=0.5\textwidth]{Figures/Holospheres/ExtinctionvsWavelength_Exploration_LRGRTXT.pdf}
    \includegraphics[page=7,width=0.5\textwidth]{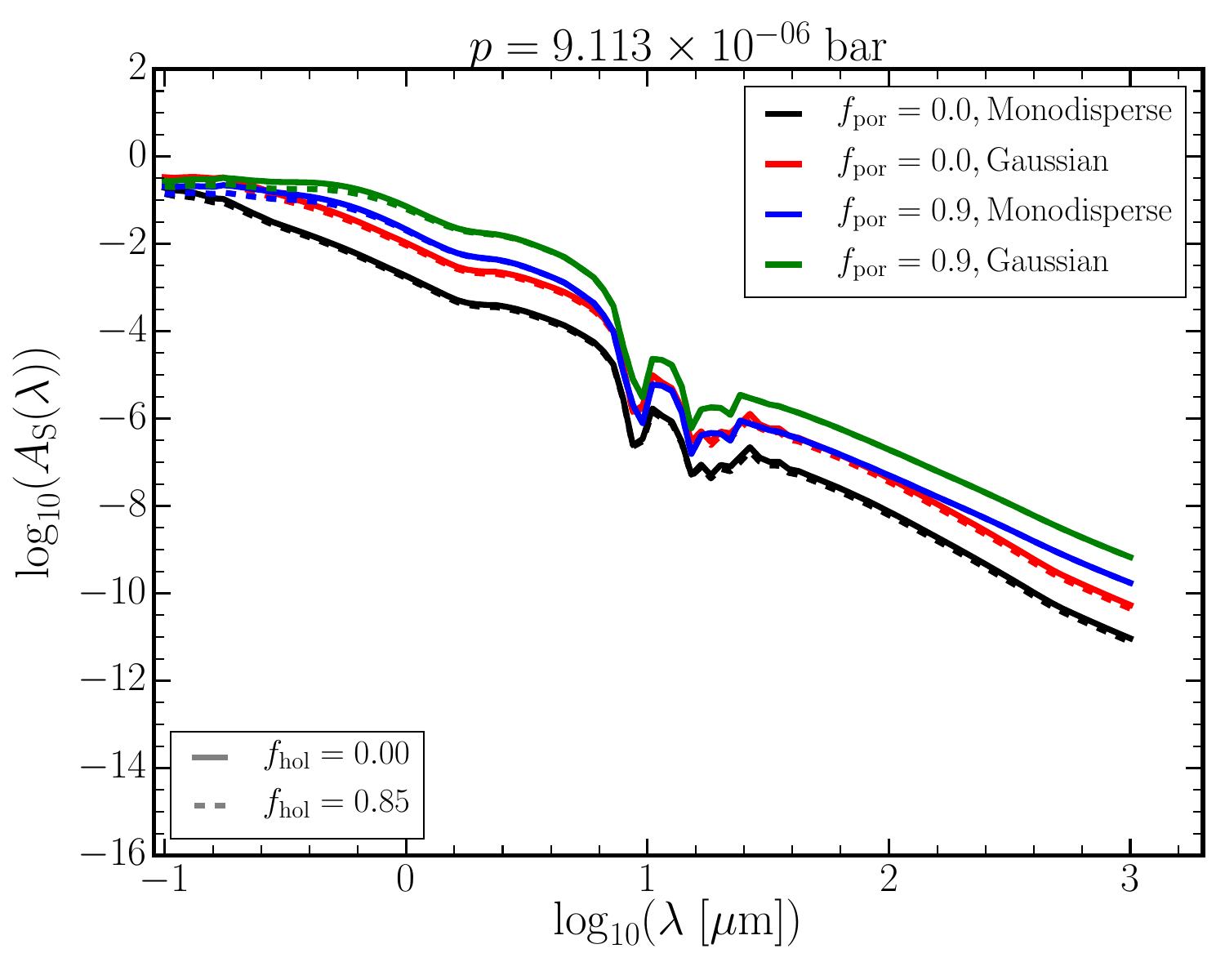}
    \includegraphics[page=7,width=0.5\textwidth]{Figures/Holospheres/ExtinctionvsWavelength_Exploration_LRGRTXT.pdf}
    \includegraphics[page=9,width=0.5\textwidth]{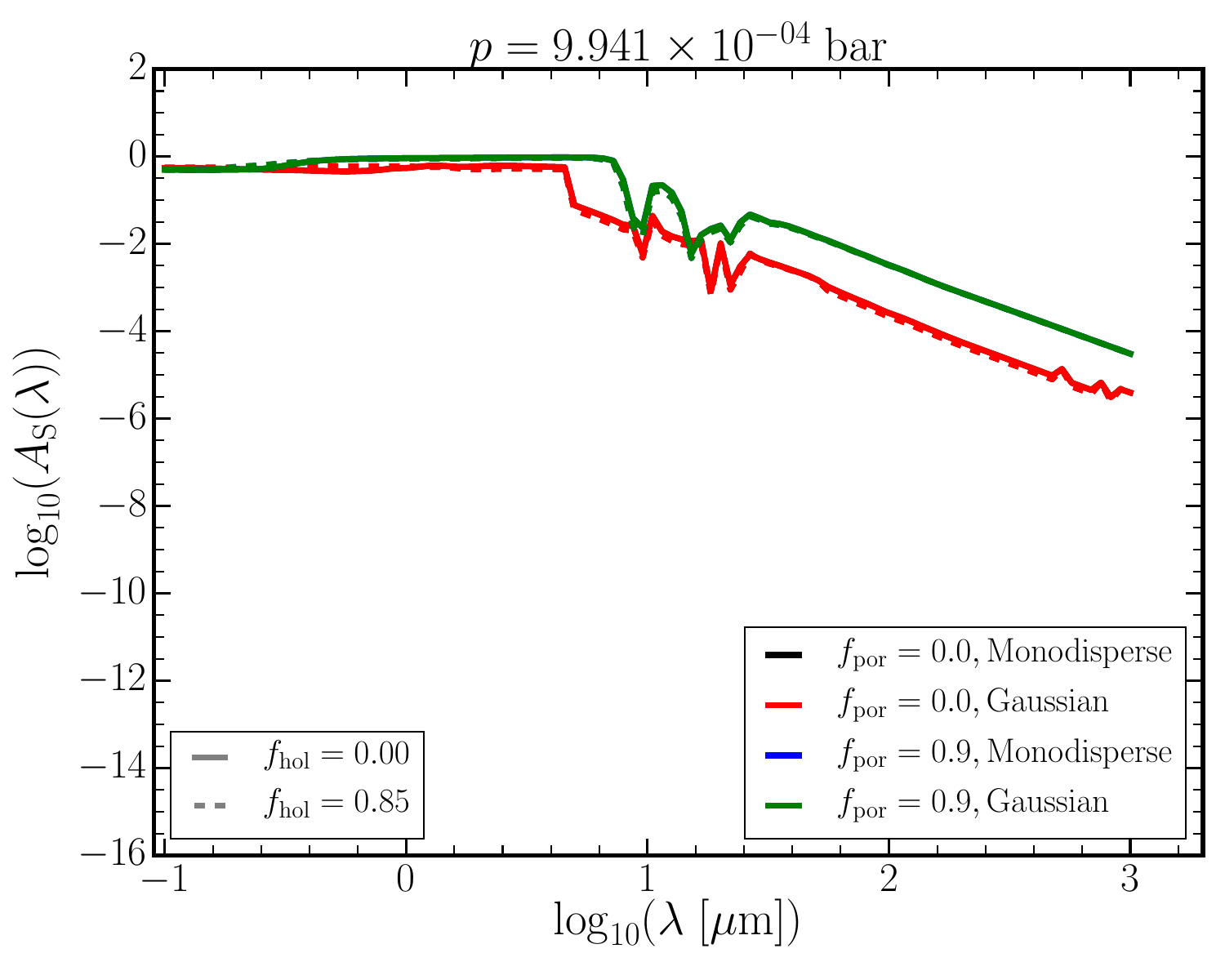}
    \includegraphics[page=9,width=0.5\textwidth]{Figures/Holospheres/ExtinctionvsWavelength_Exploration_LRGRTXT.pdf}
    \caption{Single-scattering albedo $A_{\rm S}(\lambda)$ (\textbf{left column}) and extinction cross section $C_{\rm ext}\ {\rm [cm^2]}$ (\textbf{right column}) as a function of wavelength at selected pressure levels. Calculated with (solid) and without (dashed) a distribution of hollow spheres for the $T_{\rm eff}\,=\,1800\,{\rm K},\,\log(\varg)\,=\,3.0$ profile. Shown are the pressure levels nearest to $10^{-10},\,10^{-5},\text{and}\,10^{-3}\,{\rm bar}$ (\textbf{top}, \textbf{middle}, and \textbf{bottom} panels). These levels are representative of the upper atmosphere, the typical lowest pressure of general circulation models, and the regime where transmission observations are sensitive, respectively. Four cloud particle cases are shown: monodisperse distribution of compact particles (black), Gaussian distribution of compact particles (red), monodisperse distribution of highly micro-porous particles $f_{\rm por}\,=\,0.9$ (blue), and Gaussian distribution of highly micro-porous particles (green).}
    \label{fig:holo_albedo_lambda}
\end{figure*}

\begin{figure}
    \centering
    \includegraphics[width=0.5\textwidth]{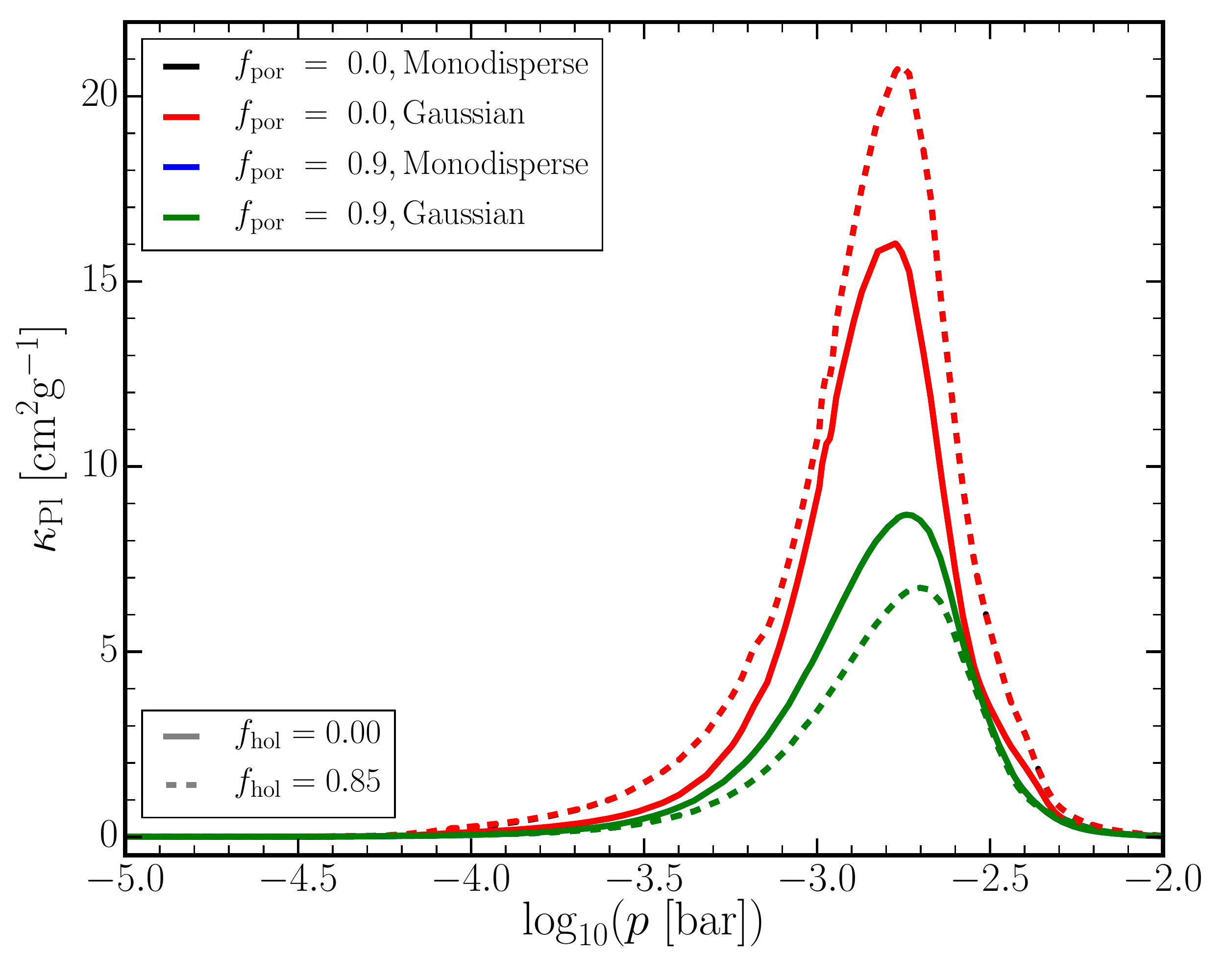}
    \caption{Planck mean opacity $\kappa_{\rm Pl}\,[\rm{cm^{^2}\,g^{-1}}]$ for the $T_{\rm eff}\,=\,1800\,{\rm K},\,\log(\varg)\,=\,3.0$ atmosphere shown from $10^{-5}$ to $10^{-2}\,{\rm bar}$. Calculated with (solid) and without (dashed) a distribution of hollow spheres. Four cloud particle cases are shown: monodisperse distribution of compact particles (black), Gaussian distribution of compact particles (red), monodisperse distribution of highly micro-porous particles $f_{\rm por}\,=\,0.9$ (blue), and Gaussian distribution of highly micro-porous particles (green).}
    \label{fig:planck_mean}
\end{figure}

Figure~\ref{fig:holo_albedo_lambda} shows the wavelength dependence of the single-scattering albedo as well as the extinction cross-section both with and without a distribution of hollow spheres for three critical pressure levels ($10^{-10},\,10^{-5},\text{and}\,10^{-3},\,{\rm bar}$). As expected, the cloud particle size distribution has little observed differences for all except $10^{-5}\,{\rm bar}$, where the cloud particle size distribution expands significantly (see Fig.~\ref{fig:Sizedist}). Notably the hollow spheres have the effect of increasing the extinction cross-section for compact particles at high pressures ($10^{-4},\,10^{-3}\,{\rm bar}$), but marginally decreasing it for the highly micro-porous case for wavelengths between $1-10\,\rm{\mu m}$. This is further highlighted by integrating across all wavelengths to determine the Planck mean opacity Fig.~\ref{fig:planck_mean}, where the peak (which aligns with the ratio of the cloud particle mass to gas mass density, see Fig.~\ref{fig:multi_porous}, bottom left) is reduced for this case. This agrees with previous work by \citet{Juncher2017}, who found a reduced Planck mean opacity for highly micro-porous particles. This implies that estimates for the mass of cloud particles in an atmosphere from retrieval methods may be significant overestimates unless the cloud particles are sufficiently porous. For the warm gas giant atmosphere, the albedo of the  cloud particles is generally very low, $\sim$~1\% for the atmosphere above $1\,{\rm mbar}$. This even holds with the enhancement of the albedo we described for the mineral snowflake clouds. This means that mineral snowflakes, like compact cloud particles, remain relatively poor reflectors and are much stronger absorbers. The extinction cross-section and single-scattering albedo both show the presence of silicate features between $10-20\,\rm{\mu m}$, which are sensitive to the assumptions on the cloud particles.

\begin{figure}
    \centering
    \includegraphics[width=0.5\textwidth]{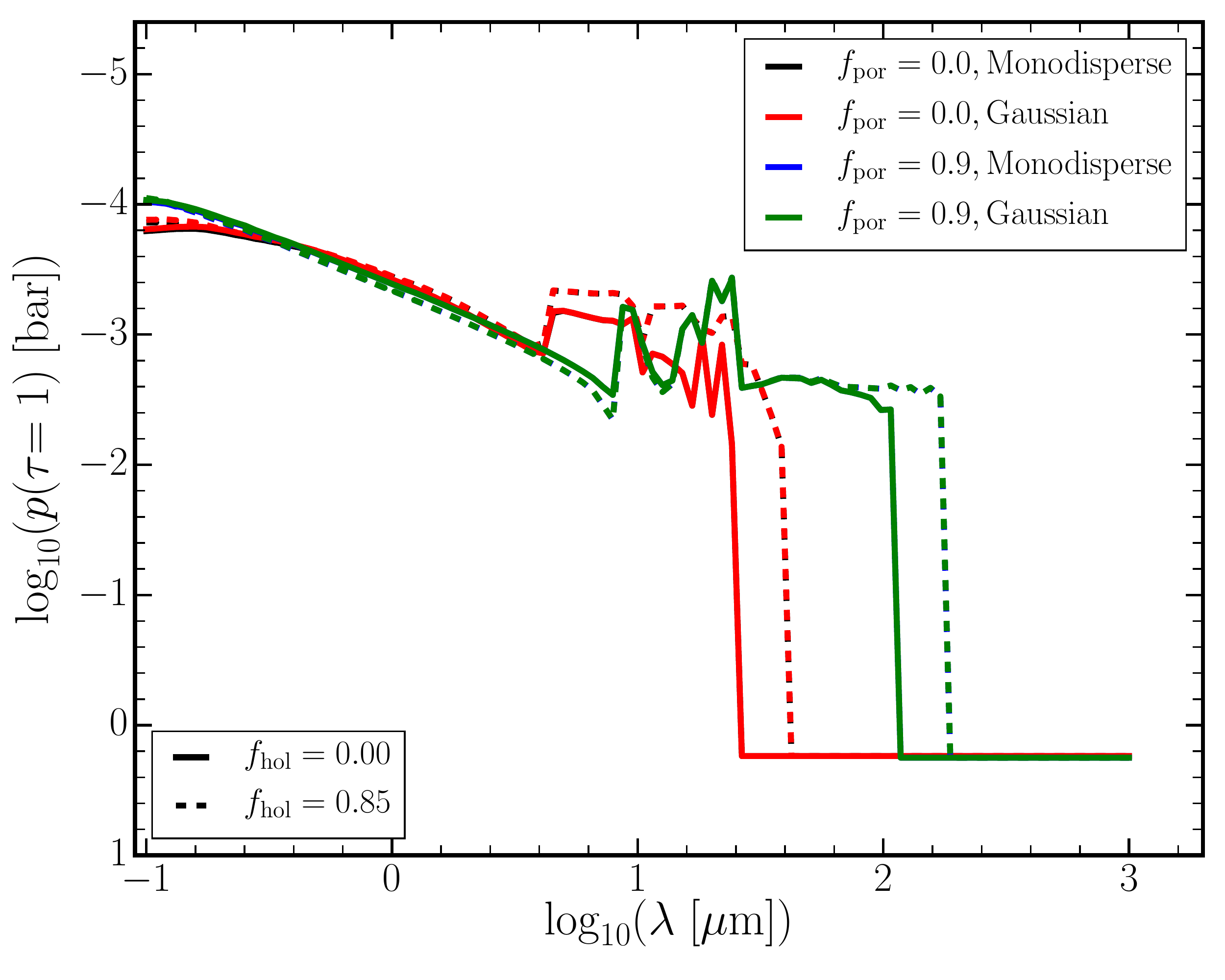}
    \caption{Atmospheric pressure level $p\,[\rm{bar}]$ at which the vertically integrated optical depth of cloud particles reaches unity ($\tau = 1$) as a function of wavelength. Four cloud particle cases are shown: monodisperse distribution of compact particles (black), Gaussian distribution of compact particles (red), monodisperse distribution of highly micro-porous particles $f_{\rm por}\,=\,0.9$ (blue), and Gaussian distribution of highly micro-porous particles (green). The y-axis is inverted because we integrate from the top of the atmosphere. The atmosphere is $T_{\rm_{eff}}\,=\,1800\,{\rm K},\,\log(\varg)\,=\,3.0$.}
    \label{fig:Opt_depth}
\end{figure}

\begin{figure}
    \centering
    \includegraphics[width=0.5\textwidth]{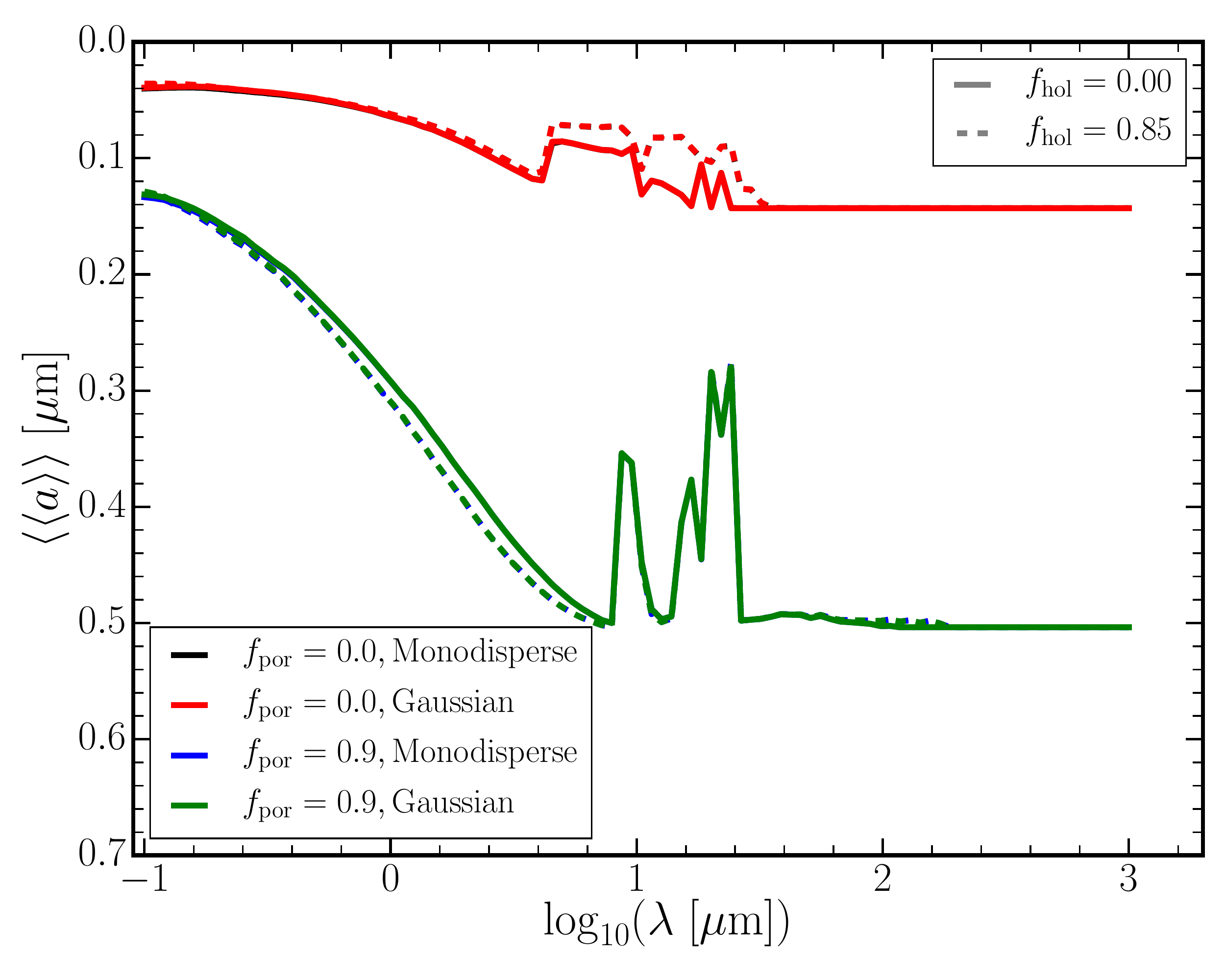}
    \caption{Average of the mean cloud particle size for particles in the optically thin atmosphere $\langle\,\langle\,a\,\rangle\,\rangle\,[\rm{\mu m}]$, (defined as the average above the pressure, as defined in Eq.~\ref{equ:ave_seen_cloudsize} and shown in Fig.~\ref{fig:Opt_depth}). Four cloud particle cases are shown: monodisperse distribution of compact particles (black), Gaussian distribution of compact particles (red), monodisperse distribution of highly micro-porous particles $f_{\rm por}\,=\,0.9$ (blue), and Gaussian distribution of highly micro-porous particles (green). The inverted y-axis aids comparison with Fig.\ref{fig:Opt_depth}.}
    \label{fig:grain_size_depth}
\end{figure}

When the optical depth of clouds is investigated using the vertically integrated pressure level (from the top of the atmosphere, TOA) at which the optical depth of the clouds reaches unity (Figure~\ref{fig:Opt_depth}), all models follow a similar slope for less than $3\,\rm{\mu m.}$   However, the micro-porous and compact cases diverge at $3\,\rm{\mu m,}$ with the compact cases exhibiting clear silicate spectral features. The highly micro-porous cases suppress these features until $\sim 8\,\rm{\mu m}$. For cases where the optical depth never reaches unity, our integration reaches the bottom of the atmosphere at $2\,{\rm bar}$. This occurs for the compact case at $\sim 18\,\rm{\mu m}$ and for the highly micro-porous cases to $> 100\,\rm{\mu m}$, which means that the depth to which {\sc JWST} {\sc MIRI} observations are sensitive (between $5\,\rm{\mu m}$ and $28\,\rm{\mu m}$ \citep{Beichman2014}) is likely an indicator of cloud micro-porosities for similar atmospheres to the warm gas giant exoplanet we considered. {\sc MIRI} observations will also include details of silicate features for similar exoplanet atmospheres. The details of these features depend on the particle micro-porosity and shape. The changes to the optical depth of the cloud particles are largely a combination of the changes to the albedo of the cloud particles and an increase in the geometrical cross-section for micro-porous particles and the larger particles in the size distribution. Hollow spheres generally also decrease the wavelength for which the clouds reach $\tau\,=\,1$ by a factor of 2 consistently for  the compact and micro-porous models. This occurs despite the reduction in albedo because of the large cross-sectional areas of cloud particles with large $f_{\rm hol}$ factors in the hollow sphere distribution. For the silicate spectral features, the optical depth reaches unity at about the millibar level, which matches the levels anticipated by observers. Retrievals deriving the mean cloud particle size \citep{Benneke2019a} will only be affected by cloud particles in the optically thin atmosphere. We investigate this in Fig.~\ref{fig:grain_size_depth} by approximating the optically thin regime as the atmosphere above the level at which $\tau\,=\,1.$ We computed the average cloud particle size across the atmosphere above this level using

\begin{equation}
        \langle\,\langle\,a\,\rangle\,\rangle\,=\,\frac{\int_{\tau = 1}^{TOA} n(a)\langle\,a\,\rangle da}{\int n(a) da}.
        \label{equ:ave_seen_cloudsize}
\end{equation}

This calculation was made in a normal geometry because our model is limited to 1D. \citet{Fortney2005} showed that a slanted geometry means that even small effects on the opacity in this regime can have massive implications for transmission spectra. As a result, the difference is a factor of 5 between the compact and highly micro-porous case, with minor deviations for inclusion of non-spherical particles. This challenges the precision of the results in \citet{Benneke2019a}, who retrieved the cloud particle size to be $0.6 \pm 0.06\,\rm{\mu m}$, but for a different atmosphere. There is no difference between the mono-disperse and Gaussian distribution because the means of the two distributions do not differ significantly (Fig.~\ref{fig:radii_comp}).

\section{Conclusion}
\label{sec:Conc}
We studied the effect of micro-porosity on the cloud structure, particle size, and material composition on exoplanets and brown dwarfs, and assessed how assumptions about cloud particles, such as sphericity, homodispersity, and compactness, affect their spectral properties. Our conclusions are listed below.

\begin{itemize}
    \item Micro-porous cloud particles (mineral snowflakes) have lower number densities and larger grain sizes than compact particles.
    \item The local material composition of cloud particles is also affected by micro-porosity, with increased bulk growth across a wider range of pressure levels.
    \item Mineral clouds are poor reflectors and are much stronger absorbers. Even with high micro-porosity, which increases the albedo over clouds made of compact particles, they remain relatively poor reflectors.
    \item Mineral snowflake clouds have an increased optical depth at near-infrared wavelengths compared with clouds of compact particles.
    \item Clouds with a wide local particle size distributions have a significantly different single-scattering albedo in the near-UV, optical, and near-infrared wavelengths compared to monodisperse clouds, which may be detectable by transmission spectroscopy.
    \item The presence of non-spherical cloud particles may be observable by a distinct cut-off wavelength for $\tau\ =\ 1$  in the {\sc JWST MIRI} bandpass.
    \item The effects of micro-porosity and non-spherical cloud particles can be most clearly separated in the spectral features of silicates between $5-20\,\rm{\mu m.}$
\end{itemize}

Other sources of porosity may generate similar or greater effects to those examined here. One source of porosity is coagulation, which is just now beginning to be examined for exoplanet atmospheres \citep{Powell2018, Kawashima2018, Ohno2018}. Coagulation will have to be examined in more detail to evaluate its contribution to observations and atmospheric retrievals.

For gas giant atmospheres, {\sc JWST MIRI} will observe at the wavelengths $5-28\,{\rm \mu m}$. In this regime (\citep{Beichman2014}), silicate spectral features of the clouds will be apparent, which will be sensitive to the details of the cloud particle model that is assumed. Furthermore, depending on the micro-porosity and shape of cloud particles, the clouds may be optically thin at the upper limits of the {\sc MIRI} wavelength range, which can be used to test assumptions about cloud particle compactness. A distribution of hollow spheres is a simple way of calculating non-spherical cloud particle effects, and has the benefit of being defined by only one additional parameter. This makes it suitable for inclusion into retrieval codes (as has been done in \cite{Molliere2019}) for cloud properties without drastically increasing the parameter space. Cloud effects on the spectra of brown dwarf and exoplanet atmospheres are a combination of the micro-porosity, non-sphericity, material composition, and cloud particle size distribution, all of which must be modelled consistently to be accurately derived. Retrieval efforts of cloud properties using simplified models must be cautiously interpreted.

\begin{acknowledgements}
D.S. acknowledges financial support from the Science and Technology Facilities Council (STFC), UK. for his PhD studentship (project reference 2093954). We thank Peter Woitke, Oliver Herbort, and Patrick Barth for their valuable discussions and support whilst writing this paper.
\end{acknowledgements}

\bibliography{Bibliography.bib}

\onecolumn{    
\begin{appendix}
\section{Additional figures and tables}
\label{sec:Apepndix}

\begin{figure}
    \centering
    \includegraphics[width=0.5\textwidth]{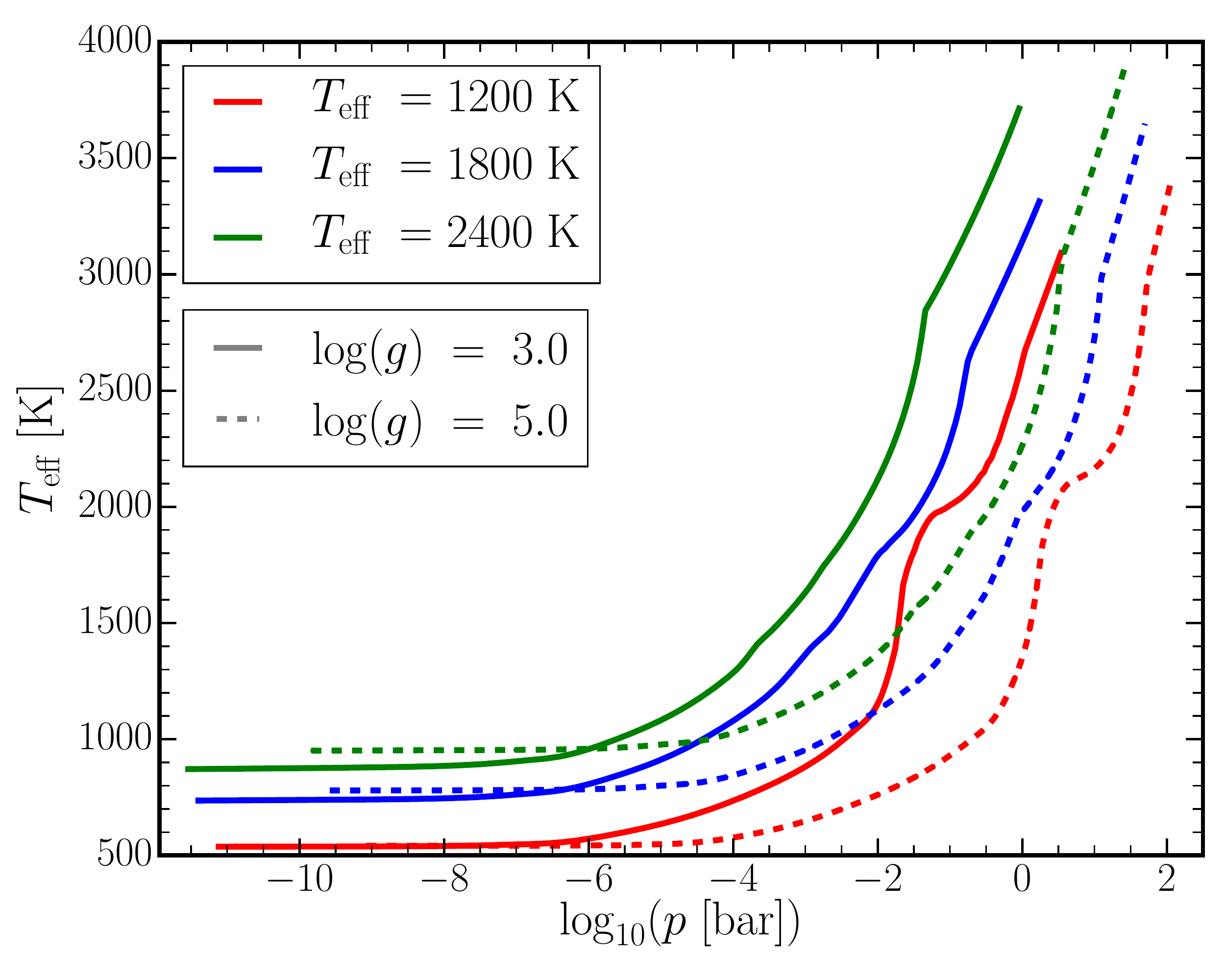}
    \caption{{\sc Drift-Phoenix} ($T_{\rm gas}$-$p_{\rm gas}$) profiles for $T_{\rm eff}\,=\,1200,\,1800,\text{and }\,2400\,{\rm K}$ (red, blue, and green, respectively) and $\log(\varg)\,=\,3.0\text{ and}\,5.0$ (solid and dashed, respectively).}
    \label{fig:DriftPhoenix}
\end{figure}

\begin{table*}[h]
  \centering
   \caption{References and wavelength coverage of optical constants for the condensate materials we considered.}
  \begin{tabular}{ccc}
     \hline\hline 
     Material Species & Reference & Wavelength Range $\rm (\mu m)$\\ \hline \hline
    \ce{TiO2}[s] (rutile) & \citet{TiO2_opt} & $0.47$--$36$\\
    \ce{SiO2}[s] (alpha-Quartz) & \citet{Palik1985}, \citet{SiO2_opt} & $0.00012$--$10000$\\
    \ce{SiO}[s] (polycrystalline)& Philipp in \citet{Palik1985} & $0.0015$--$14$\\
    \ce{MgSiO3}[s] (glass)& \citet{MgSiO3_opt} & $0.20$--$500$\\
    \ce{Mg2SiO4}[s] (crystalline)& \citet{Suto2006} & $0.10$--$1000$\\
    \ce{MgO}[s] (cubic)& \citet{Palik1985} & $0.017$--$625$\\
    \ce{Fe}[s] (metallic)& \citet{Palik1985} & $0.00012$--$285$\\
    \ce{FeO}[s] (amorphous)& \citet{FeO_opt} & $0.20$--$500$\\
    \ce{Fe2O3}[s] (amorphous)& Amaury H.M.J. Triaud (priv. comm.) & $0.10$--$1000$\\
    \ce{Fe2SiO4}[s] (amorphous)& \citet{MgSiO3_opt} & $0.20$--$500$ \\
    \ce{FeS}[s] (amorphous)& Henning (unpublished) & $0.10$--$100000$\\
    \ce{CaTiO3}[s] (amorphous)& \citet{Posch2003} & $2$--$5843$\\
    \ce{CaSiO3}[s] & No data - treated as vacuum & N/A\\
    \ce{Al2O3}[s] (glass)& \citet{Begemann1997} & $0.10$--$200$ \\
    \ce{C}[s] (graphite)& \citet{Palik1985} & $0.20$--$794$\\\hline\hline
  \end{tabular}
  \label{tab:opt}
\end{table*}

\begin{figure*}
    \centering
    \includegraphics[width=0.33\textwidth,page=1]{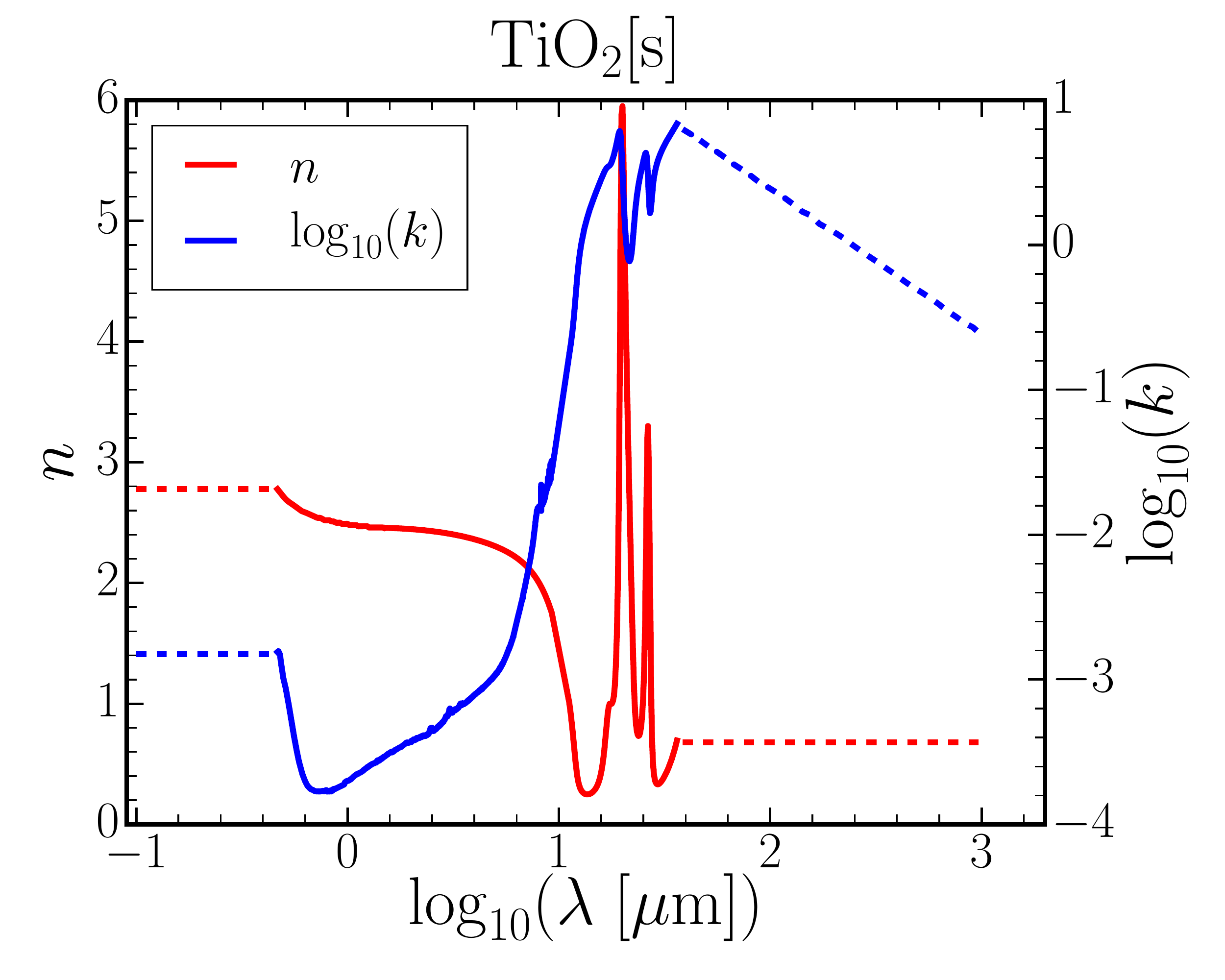}
    \includegraphics[width=0.33\textwidth,page=5]{Figures/Appendix/SW_Material_Refractive_Indexes_APPENDIX.pdf}
    \includegraphics[width=0.33\textwidth,page=4]{Figures/Appendix/SW_Material_Refractive_Indexes_APPENDIX.pdf}
    \includegraphics[width=0.33\textwidth,page=14]{Figures/Appendix/SW_Material_Refractive_Indexes_APPENDIX.pdf}
    \includegraphics[width=0.33\textwidth,page=2]{Figures/Appendix/SW_Material_Refractive_Indexes_APPENDIX.pdf}
    \includegraphics[width=0.33\textwidth,page=13]{Figures/Appendix/SW_Material_Refractive_Indexes_APPENDIX.pdf}
    \includegraphics[width=0.33\textwidth,page=6]{Figures/Appendix/SW_Material_Refractive_Indexes_APPENDIX.pdf}
    \includegraphics[width=0.33\textwidth,page=10]{Figures/Appendix/SW_Material_Refractive_Indexes_APPENDIX.pdf}
    \label{fig:appendix_refindex1}
    \includegraphics[width=0.33\textwidth,page=12]{Figures/Appendix/SW_Material_Refractive_Indexes_APPENDIX.pdf}
    \includegraphics[width=0.33\textwidth,page=15]{Figures/Appendix/SW_Material_Refractive_Indexes_APPENDIX.pdf}
    \includegraphics[width=0.33\textwidth,page=11]{Figures/Appendix/SW_Material_Refractive_Indexes_APPENDIX.pdf}
    \includegraphics[width=0.33\textwidth,page=8]{Figures/Appendix/SW_Material_Refractive_Indexes_APPENDIX.pdf}
    \includegraphics[width=0.33\textwidth,page=7]{Figures/Appendix/SW_Material_Refractive_Indexes_APPENDIX.pdf}
    \includegraphics[width=0.33\textwidth,page=16]{Figures/Appendix/SW_Material_Refractive_Indexes_APPENDIX.pdf}
    \caption{Real and imaginary refractive indexes (red and blue, respectively) for the condensate materials we used, shown across the wavelength range $0.1-1000\,{\rm \mu m}$. Solid lines indicate regions for which reference data exist, and dashed lines indicate regions that were extrapolated. References for the refractive index data are found in Table \ref{tab:opt}.}
    \label{fig:appendix_refindex2}
\end{figure*}

\begin{figure}
    \includegraphics[width=0.5\textwidth,page=1]{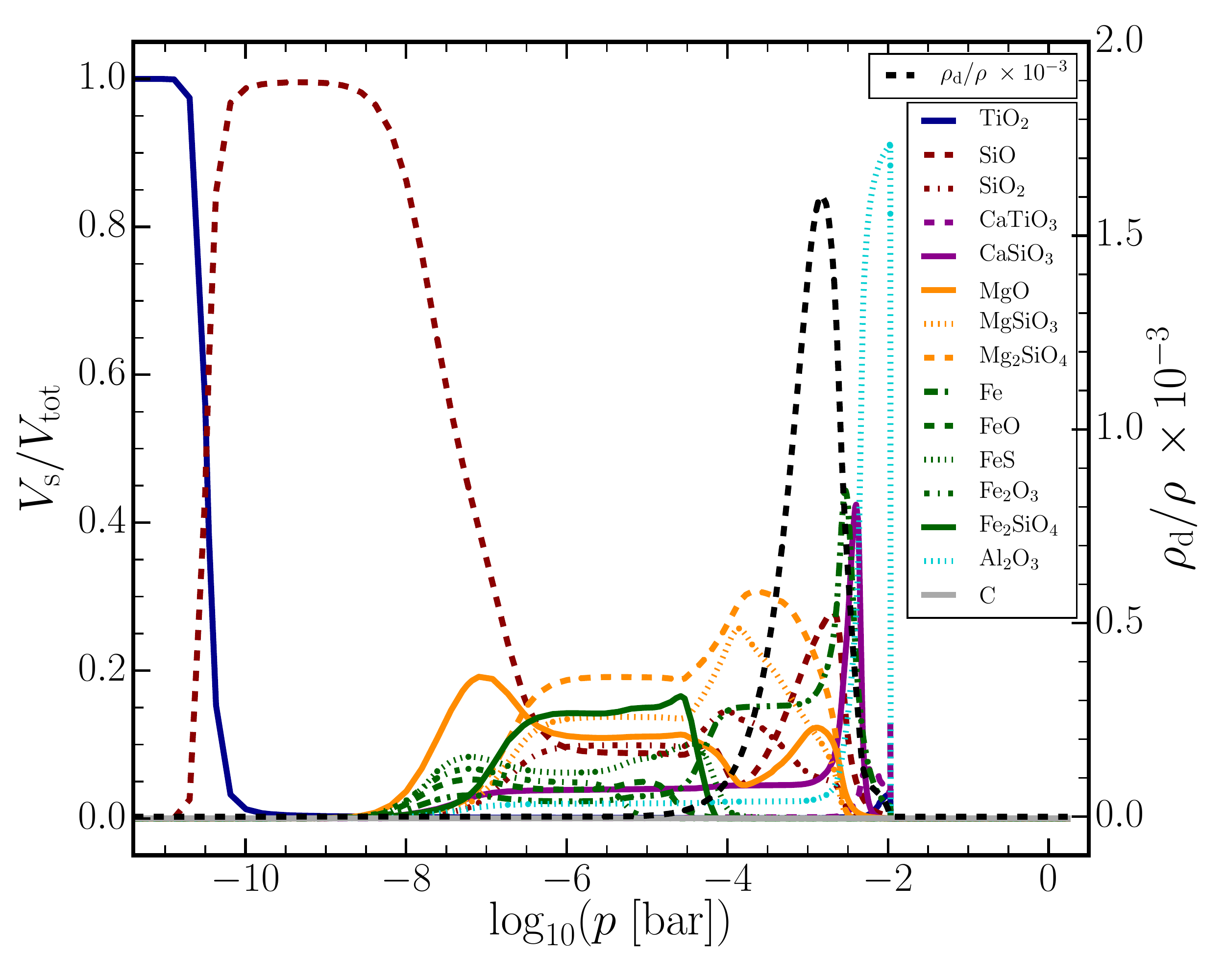}
    \includegraphics[width=0.5\textwidth,page=5]{Figures/Appendix/Material_Comp_Plots_LRGRTXT.pdf}
    \caption{Material properties of cloud particles in the $T_{\rm_{eff}}\,=\,1800\,{\rm K},\,\log(\varg)\,=\,3.0$ atmosphere. \textbf{Left:} For compact cloud particles (f$_{\rm por}\,=\,0.0$). \textbf{Right:} For highly micro-porous cloud particles (f$_{\rm por}\,=\,0.9$). \textbf{Left axis:} Material volume fractions of cloud particles. \textbf{Right axis:} Ratio of cloud particle mass to gas mass $\rho_{\rm d}/\rho$ scaled by a factor of $10^{-3}$ (black dashed line).}
    \label{fig:mat_comp}
\end{figure}

\begin{figure}
    \centering
    \includegraphics[width=0.5\textwidth]{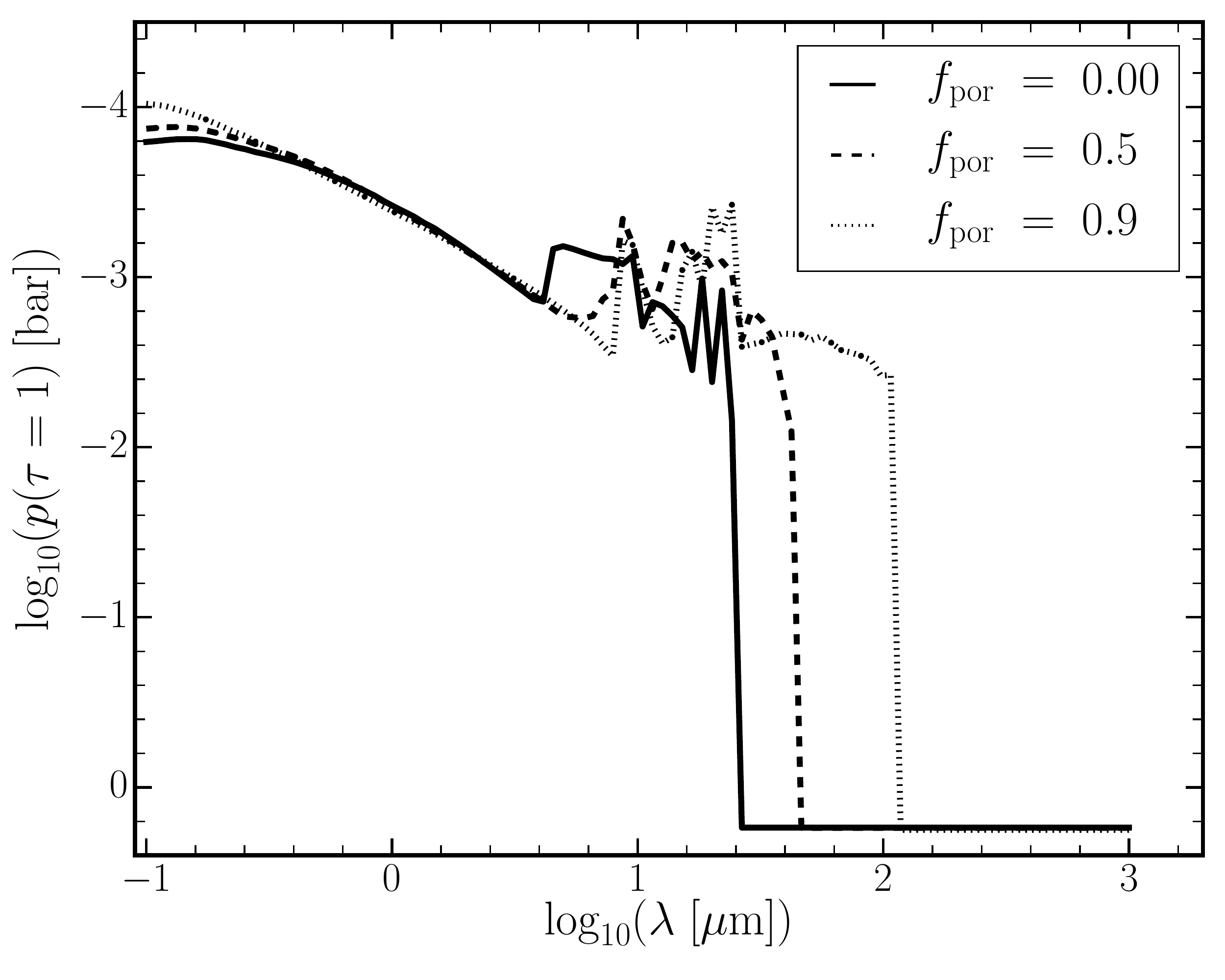}
    \caption{Pressure level at which the cloud deck optical depth reaches unity ($p(\tau\,=\,1)$) for wavelengths in the range $\lambda = 0.1-1000\,\rm{\mu m}$ integrated from the top of the atmosphere.}
    \label{fig:porous_opacity}
\end{figure}

\begin{figure}
        \centering
        \includegraphics[width=0.49\textwidth,page=1]{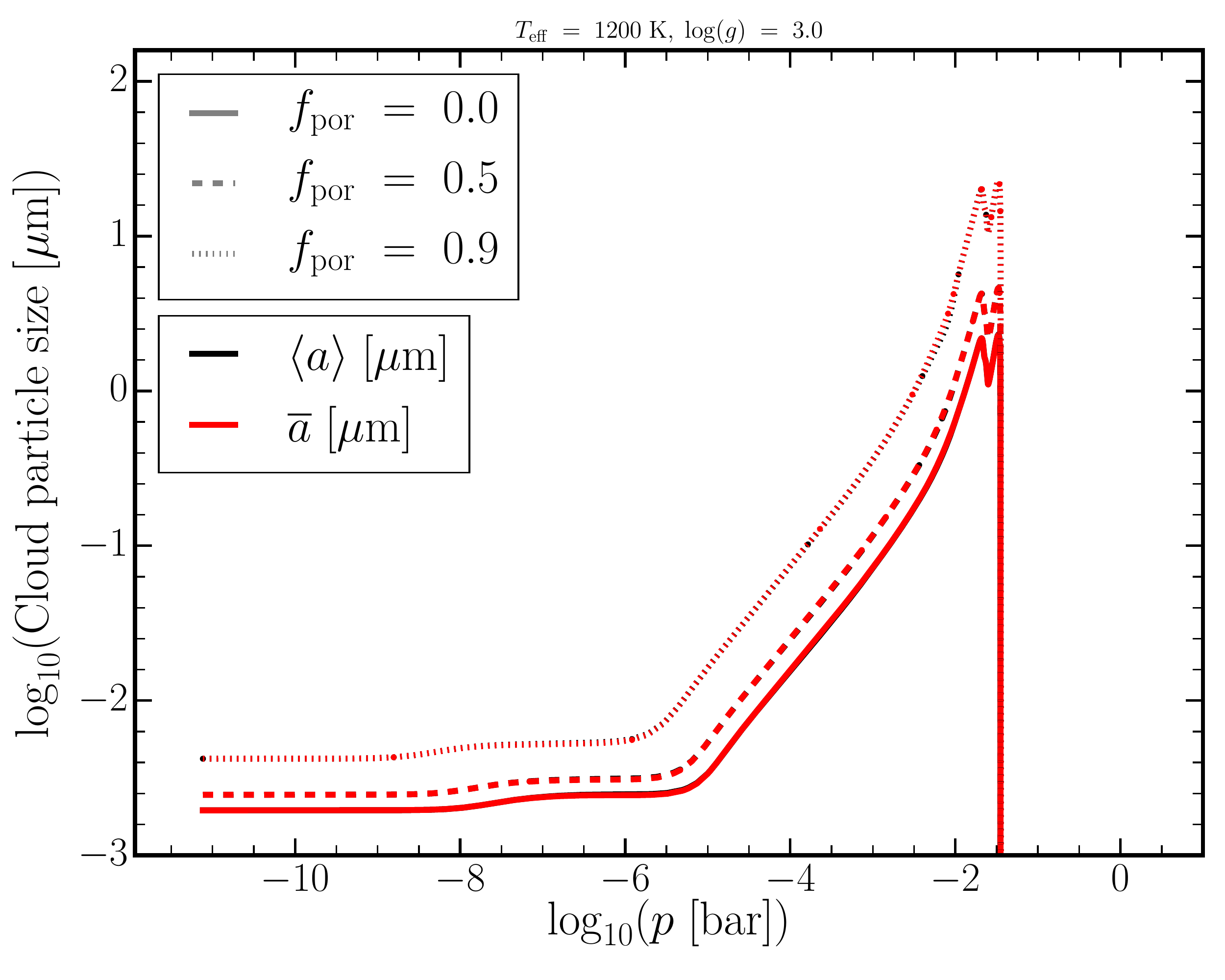}
        \includegraphics[width=0.49\textwidth,page=2]{Figures/Appendix/Radii_Comparison_LRGRTXT.pdf}
        \caption{Comparison of the two methods of calculating the cloud particle size, the mean particle size from the moments ($\langle\,a\,\rangle$) according to Eq.~\ref{equ:meansize}, and from the size distribution ($\overline{a}$) using Eq.~\ref{equ:gau_grainrad}. \textbf{Left:} $\langle\,a\,\rangle$ and $\overline{a}$ throughout the atmosphere for $T_{\rm eff}\,=\,1200\,{\rm K},\,\log(\varg)\,=\,3.0$, and three micro-porosity cases $f_{\rm por}\,=\,0.0,\,0.5,\text{and}\,0.9$. \textbf{Right:} $\overline{a}$ as a fraction of $\langle\,a\,\rangle$ for the same atmosphere and micro-porosity cases as above. We chose this atmosphere because it shows the largest difference between the two average cloud particle sizes across all profiles.}
    \label{fig:radii_comp}
\end{figure}
\end{appendix}
}
\end{document}